\documentclass[10pt,twocolumn,prl,aps,showpacs,superscriptaddress,floatfix,preprintnumbers,reprint,nofootinbib]{revtex4-1}
\usepackage{color}
\usepackage{epsfig}
\usepackage{amsmath,bm}
\usepackage{subcaption}
\usepackage{blindtext}

\newcommand{\sub}[2]{\ensuremath{#1_{\mathrm{#2}}}}

\newcommand{\like}[2]{\ensuremath{\mathcal{L}_{\mathrm{#1}}\left(#2\right)}}

\newcommand{\likeone}[1]{\ensuremath{\mathcal{L}_{\mathrm{#1}}}}

\newcommand{\boldvec}[1]{\ensuremath{\boldsymbol{#1}}}
\newcommand{\unit}[2]{#1 \; \mathrm{#2}}

\newcommand{\code}[1]{\textalltt{#1}}
      \makeatletter
      \newcommand*{\textalltt}{}
      \DeclareRobustCommand*{\textalltt}{%
	      \begingroup
	      \let\do\@makeother
	      \dospecials
	      \catcode`\\=\z@
	      \catcode`\{=\@ne
	      \catcode`\}=\tw@
	      \verbatim@font\@noligs
	      \@vobeyspaces
	      \frenchspacing
	      \@textalltt
      }
      \newcommand*{\@textalltt}[1]{%
	      #1%
	      \endgroup
      }
      \makeatother
      
\newcommand{\figref}[1]{fig.~\ref{#1}}
\newcommand{\figThreeref}[3]{figs.~\ref{#1},~\ref{#2},~\ref{#3}}

\newcommand{\Tabref}[1]{Tab.~\ref{#1}}

\begin{document}
%%%%%%%%%%%%%%%%%%%%%%%%%%%%%%%%%%%%%%%%%%%%%%%%%%%%%%%%%%%%%%%%%%%%%%%%%%%%%%%
% </Document definition and begin>
%%%%%%%%%%%%%%%%%%%%%%%%%%%%%%%%%%%%%%%%%%%%%%%%%%%%%%%%%%%%%%%%%%%%%%%%%%%%%%%

%%%%%%%%%%%%%%%%%%%%%%%%%%%%%%%%%%%%%%%%%%%%%%%%%%%%%%%%%%%%%%%%%%%%%%%%%%%%%%%
% <Title and authors>
%%%%%%%%%%%%%%%%%%%%%%%%%%%%%%%%%%%%%%%%%%%%%%%%%%%%%%%%%%%%%%%%%%%%%%%%%%%%%%%
\title{A Global Bayesian Analysis of Neutrino Mass Data}
\preprint{MPP-2017-79}

\author{Allen Caldwell}\email{caldwell@mpp.mpg.de}
\affiliation{Max-Planck-Institut f\"ur Physik
(Werner-Heisenberg-Institut), F\"ohringer Ring 6, 80805 M\"unchen,
Germany}

\author{Manuel Ettengruber}\email{manuel@mpp.mpg.de}
\affiliation{Max-Planck-Institut f\"ur Physik
(Werner-Heisenberg-Institut), F\"ohringer Ring 6, 80805 M\"unchen,
Germany}

\author{Alexander Merle}\email{amerle@mpp.mpg.de}
\affiliation{Max-Planck-Institut f\"ur Physik
(Werner-Heisenberg-Institut), F\"ohringer Ring 6, 80805 M\"unchen,
Germany}

\author{Oliver Schulz}\email{oschulz@mpp.mpg.de}
\affiliation{Max-Planck-Institut f\"ur Physik
(Werner-Heisenberg-Institut), F\"ohringer Ring 6, 80805 M\"unchen,
Germany}

\author{Maximilian Totzauer}\email{totzauer@mpp.mpg.de}
\affiliation{Max-Planck-Institut f\"ur Physik
(Werner-Heisenberg-Institut), F\"ohringer Ring 6, 80805 M\"unchen,
Germany}

\date{\today}
%%%%%%%%%%%%%%%%%%%%%%%%%%%%%%%%%%%%%%%%%%%%%%%%%%%%%%%%%%%%%%%%%%%%%%%%%%%%%%%
% </Title and authors>
%%%%%%%%%%%%%%%%%%%%%%%%%%%%%%%%%%%%%%%%%%%%%%%%%%%%%%%%%%%%%%%%%%%%%%%%%%%%%%%

%%%%%%%%%%%%%%%%%%%%%%%%%%%%%%%%%%%%%%%%%%%%%%%%%%%%%%%%%%%%%%%%%%%%%%%%%%%%%%%
% <Abstract>
%%%%%%%%%%%%%%%%%%%%%%%%%%%%%%%%%%%%%%%%%%%%%%%%%%%%%%%%%%%%%%%%%%%%%%%%%%%%%%%
\begin{abstract}
We perform a global Bayesian analysis of currently available neutrino data, putting data from oscillation experiments, neutrinoless double beta decay ($0\nu\beta\beta$), and precision cosmology on an equal footing. 
%The Bayesian framework allows us to be explicit in our assumptions: 
We evaluate the discovery potential of future $0\nu\beta\beta$ experiments and the Bayes factor of the two possible neutrino mass ordering schemes for different prior choices. We show that the indication for normal ordering is still very mild and does not strongly depend on realistic prior assumptions or different combinations of cosmological data sets. 
We find a wide range for $0\nu\beta\beta$ discovery potential, depending on the absolute neutrino mass scale,
mass ordering and achievable background level. \\
\textit{This is an updated version of the original paper dated 4 May 2017. The update contains a correction to an error found in our use of the PLANCK collaboration data. All relevant figures were updated but no new data were included.}
%Our approach is a very valuable contribution to the global picture of cutting-edge neutrino physics. In particular, it can easily incorporate future data as it becomes available.
\end{abstract}
%%%%%%%%%%%%%%%%%%%%%%%%%%%%%%%%%%%%%%%%%%%%%%%%%%%%%%%%%%%%%%%%%%%%%%%%%%%%%%%
% </Abstract>
%%%%%%%%%%%%%%%%%%%%%%%%%%%%%%%%%%%%%%%%%%%%%%%%%%%%%%%%%%%%%%%%%%%%%%%%%%%%%%%

%%%%%%%%%%%%%%%%%%%%%%%%%%%%%%%%%%%%%%%%%%%%%%%%%%%%%%%%%%%%%%%%%%%%%%%%%%%%%%%
% <maketitle>
%%%%%%%%%%%%%%%%%%%%%%%%%%%%%%%%%%%%%%%%%%%%%%%%%%%%%%%%%%%%%%%%%%%%%%%%%%%%%%%
\maketitle
%%%%%%%%%%%%%%%%%%%%%%%%%%%%%%%%%%%%%%%%%%%%%%%%%%%%%%%%%%%%%%%%%%%%%%%%%%%%%%%
% </matketitle>
%%%%%%%%%%%%%%%%%%%%%%%%%%%%%%%%%%%%%%%%%%%%%%%%%%%%%%%%%%%%%%%%%%%%%%%%%%%%%%%

%%%%%%%%%%%%%%%%%%%%%%%%%%%%%%%%%%%%%%%%%%%%%%%%%%%%%%%%%%%%%%%%%%%%%%%%%%%%%%%
% <Introduction>
%%%%%%%%%%%%%%%%%%%%%%%%%%%%%%%%%%%%%%%%%%%%%%%%%%%%%%%%%%%%%%%%%%%%%%%%%%
\section{\label{sec:Intro}Introduction}
%%%%%%%%%%%%%%%%%%%%%%%%%%%%%%%%%%%%%%%%%%%%%%%%%%%%%%%%%%%%%%%%%%%%%%%%%%

Neutrino physics is one of the most attractive fields to look for new physics, and many parameters in the neutrino sector are not yet determined. A series of oscillation experiments has established the fact that at least two neutrino masses are distinct from zero, but their smallness cannot be accommodated within the Standard Model of Particle Physics. Furthermore, quantities like the Dirac phase \sub{\delta}{CP} describing the difference between matter and antimatter, the absolute neutrino mass scale, or the mass ordering (i.e., which neutrino is the lightest) are currently only poorly restricted~\cite{Esteban:2016qun}. The even more fundamental question of whether neutrinos are of Majorana or of Dirac nature (i.e., whether they are identical to their antiparticles, or not) is also still unanswered. A global analysis, combining all relevant experimental results and using the current data as efficiently as possible, is the most suitable way to address these open issues. It additionally informs upcoming experimental choices.

The first fully comprehensive Bayesian analysis of this kind is presented in this letter. We use information from oscillation experiments, precision cosmology, and neutrinoless double beta decay ($0\nu\beta\beta$). While previous works focus mostly on one of these aspects, we adopt a fully global view. Our analysis is based on the minimal framework of three light Majorana neutrinos, the most predictive setting for neutrino physics. We use global oscillation data from the \code{nu-fit} collaboration~\cite{Esteban:2016qun}, cosmological data from the Planck Legacy Archive (PLA),\footnote{Based on observations by Planck (\url{http://www.esa.int/Planck}).} and data from the $0\nu\beta\beta$-experiments KamLAND-Zen~\cite{KamLAND-Zen:2016pfg}, EXO-200~\cite{Albert:2014awa}, and GERDA~\cite{Agostini:2017iyd}. We do not include single $\beta$ decay results as they currently do not provide additional constraints.

We extract the implications for future $0\nu\beta\beta$-experiments and also address whether current data may already exhibit a tendency towards the normal ordering scheme -- which is currently under intense debate~\cite{Gerbino:2016ehw,Capozzi:2017ipn,Simpson:2017qvj,Schwetz:2017fey}.

Bayes' Theorem provides the logical path from the probability of data under different hypotheses $H_i$, $P\left(D|H_i\right)$, to the probability of a hypothesis being correct given the data. It requires the explicit definition of prior probabilities. Any other approach is either incoherent (for a discussion of $p$-values as evidence, see e.g.~\cite{Schervish:PValues}), requires predefined error rates (Neyman frequentist approach~\cite{Neyman1977-NEYFPA}), or contains implicit prior choices -- all of which are to be disfavored. If the outcome depends strongly on the choice of prior, this is simply an indication that the data is not powerful enough to draw a conclusion. In this context, it is important to find the implications from a range of reasonable prior choices.

For our analysis we assign equal prior probabilities to the two mass orderings. For the absolute neutrino mass scale, the choice of the prior on the mass, $m_{\rm lightest}$, of the lightest neutrino is critical. At present, only an upper limit is available.  We consider two choices: A prior flat in $m_{\rm lightest}$ (this will tend to favor larger values and reflects the sizes of the measured solar and atmospheric mass square differences $\Delta m^2_\odot \equiv m_2^2 - m_1^2$ and $\Delta m^2_A \equiv |m_3^2 - m_1^2|$) and a scale-free prior (flat in the logarithm of $m_{\rm lightest}$, which tends to favor small values and reflects the fact that the absolute neutrino mass scale is still unknown).  We then evaluate the consequences of both prior choices on the posterior probabilities.

Technical details regarding priors and likelihoods are provided in the supplementary material.

%%%%%%%%%%%%%%%%%%%%%%%%%%%%%%%%%%%%%%%%%%%%%%%%%%%%%%%%%%%%%%%%%%%%%%%%%%
\section{\label{sec:ParamSpaceLikePriors}Parameter Space, Likelihoods, Priors}
%%%%%%%%%%%%%%%%%%%%%%%%%%%%%%%%%%%%%%%%%%%%%%%%%%%%%%%%%%%%%%%%%%%%%%%%%%

%%%%%%%%%%%%%%%%%%%%%%%%%%%%%%%%%%%%%%%%%%%%%%%%%%%%%%%%%%%%%%%%%%%%%%%%%%
\paragraph{Parameter space}
%%%%%%%%%%%%%%%%%%%%%%%%%%%%%%%%%%%%%%%%%%%%%%%%%%%%%%%%%%%%%%%%%%%%%%%%%%

Our analysis depends on eight parameters, aggregated into the vector $\boldvec{\theta}$:
\begin{align}
 \boldvec{\theta}=\left(\sub{m}{lightest}, \Delta m^2_\odot, \Delta m^2_A, s^2_{12}, s^2_{13}, \alpha_1, \alpha_2, \mathcal{G}\right) \,.
 \label{eq:Def:ParamVector}
\end{align}
Here, $\sub{m}{lightest}$ is the smallest neutrino mass eigenvalue, while $s^2_{ij}$ denotes the sine squared of the mixing angle $\theta_{ij}$. The Majorana phases $\alpha_1$ and $\alpha_2$ take trivial values if neutrinos are of Dirac nature. The nuclear matrix elements (NMEs) required to calculate rates of $0\nu\beta\beta$ for different considered isotopes are condensed into $\mathcal{G}$. Note that $\boldvec{\theta}$ does \emph{not} contain the Dirac phase $\sub{\delta}{CP}$ or $s^2_{23}$, as they only affect neutrinos oscillations, but neither $0\nu\beta\beta$ nor cosmology. Furthermore we allow for the choice of either normal (NO: $m_1 < m_2 < m_3$) or inverted (IO: $m_3 < m_1 < m_2$) ordering.\footnote{Such an ``external'' parameter is called \emph{hyperparameter}.}

%%%%%%%%%%%%%%%%%%%%%%%%%%%%%%%%%%%%%%%%%%%%%%%%%%%%%%%%%%%%%%%%%%%%%%%%%%
\paragraph{Likelihood}
%%%%%%%%%%%%%%%%%%%%%%%%%%%%%%%%%%%%%%%%%%%%%%%%%%%%%%%%%%%%%%%%%%%%%%%%%%

\begin{figure*}[t]
\begin{subfigure}[c]{0.4\textwidth}
  \includegraphics[width=1.0\textwidth]{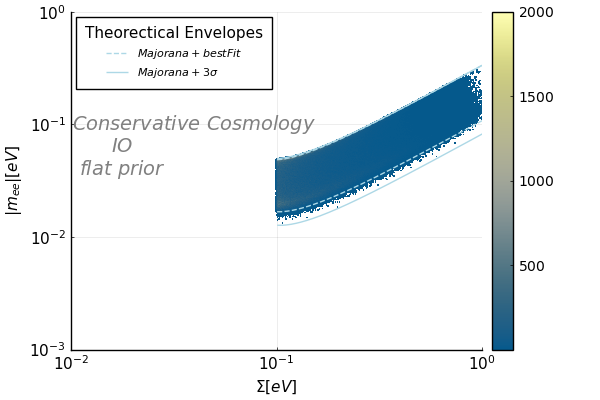}
\end{subfigure}
\begin{subfigure}[c]{0.4\textwidth}
  \includegraphics[width=1.0\textwidth]{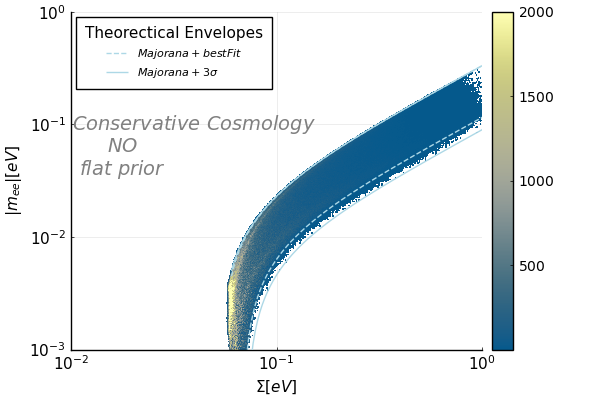}
\end{subfigure}
\\
\begin{subfigure}[c]{0.4\textwidth}
  \includegraphics[width=1.0\textwidth]{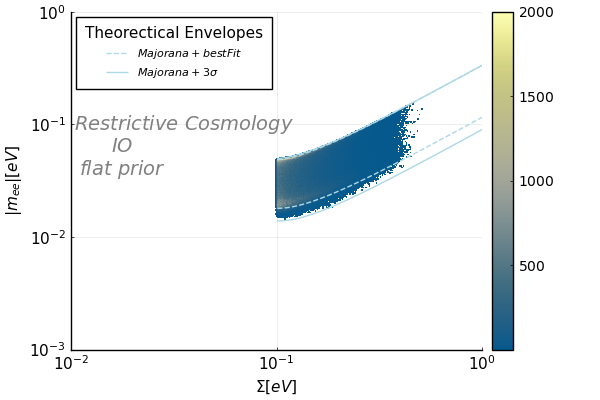}
\end{subfigure}
\begin{subfigure}[c]{0.4\textwidth}
  \includegraphics[width=1.0\textwidth]{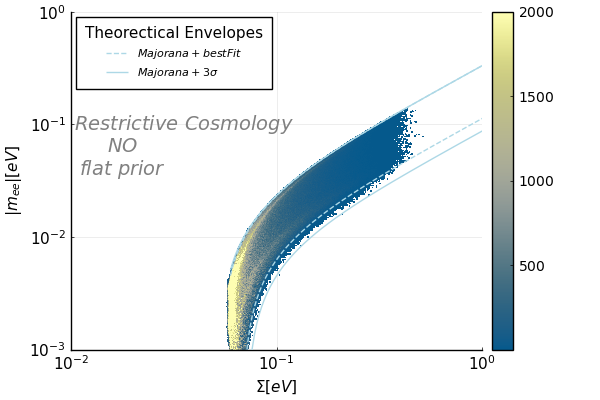}
\end{subfigure}
\\
\begin{subfigure}[c]{0.4\textwidth}
  \includegraphics[width=1.0\textwidth]{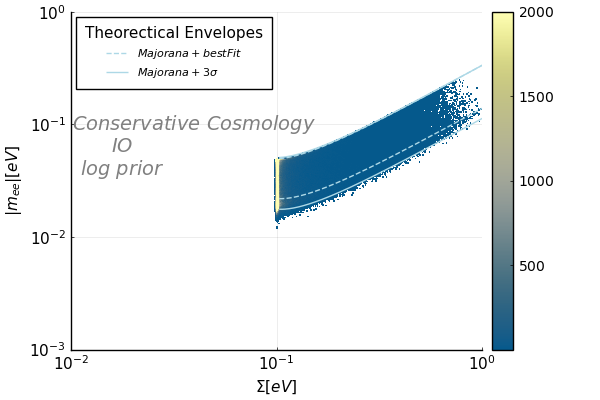}
\end{subfigure}
\begin{subfigure}[c]{0.4\textwidth}
  \includegraphics[width=1.0\textwidth]{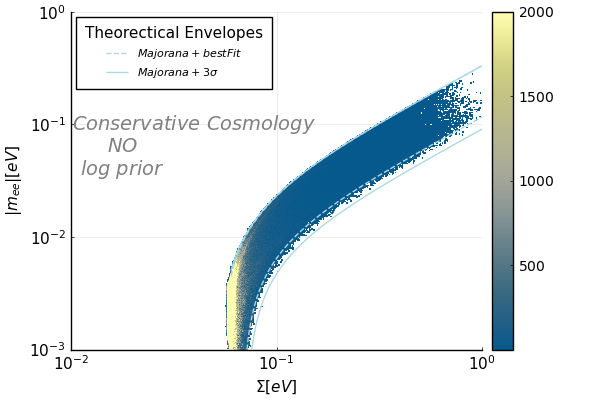}
\end{subfigure}
\\
\begin{subfigure}[c]{0.4\textwidth}
  \includegraphics[width=1.0\textwidth]{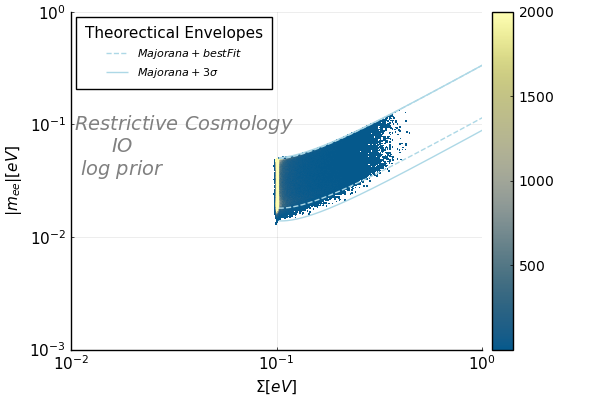}
\end{subfigure}
\begin{subfigure}[c]{0.4\textwidth}
  \includegraphics[width=1.0\textwidth]{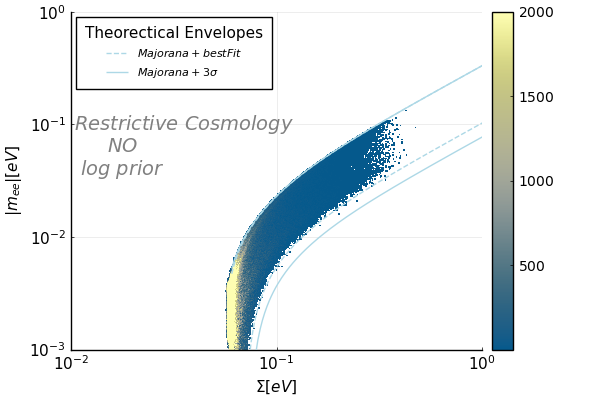}
\end{subfigure}
\caption{\label{fig:meeSigma}Heat map of posterior probability density for both combinations of cosmological data sets and choices of the prior on $\sub{m}{lightest}$. The upper panels depict the flat prior while the lower panels show the log prior. See text for definitions of the different quantities.}
\end{figure*}

The global likelihood function \likeone{glob} is a function that depends non-trivially on all eight parameters. Since every suite of experiment constitutes an independent data set, we can factorize the likelihood:
\begin{align}
  \likeone{glob} = \likeone{cosmo} \times \likeone{0\nu\beta\beta} \times \likeone{osc} \;.
\label{eq:LikeFactors}
\end{align}
The cosmological term $\likeone{cosmo}$ only depends on the sum of neutrino masses,
\begin{align}
 \Sigma  \equiv m_1 + m_2 + m_3 \;,
 \label{eq:DefSigma}
\end{align}
and hence on $\sub{m}{lightest}$, $\Delta m^2_\odot$, and $\Delta m^2_A$. We reconstructed the likelihood from the Markov chains that are publicly available on the PLA. As those chains were sampled on a prior flat in $\Sigma$, one can construct a function of $\Sigma$ that is directly proportional to $\likeone{cosmo}$ from these chains.\footnote{Note that the choice of prior used by the Planck collaboration is, in fact, unphysical, since $\Sigma$ has a minimal value $\Sigma_{\rm min}\neq 0$. As shown in~\cite{Gerbino:2016ehw}, using a more realistic prior, the resulting constraints become stronger, but they constrain $(\Sigma - \Sigma_{\rm min})$, making the corresponding limit on $\Sigma$ appear looser.} To be agnostic about the choice of datasets used in our analysis, we chose two models: both contain the $TT$ correlation of the Cosmic Microwave Background (CMB), a joint likelihood on $TT$, $EE$, $BB$, and $TE$ correlations\footnote{Here, $T$ refers to the temperature of the CMB, while $E$ and $B$ denote the basic patterns of its polarization.} for low multipoles ($2\leq l \leq29$), as well as weak gravitational lensing data. We refer to this combination of data as \emph{conservative model}. Augmenting the data by data from Baryonic Accoustic Oscillations (BAO) defines a \emph{restrictive model}, which puts tighter constraints onto $\Sigma$.\footnote{We use the PLA data sets \code{base\_mnu\_plikHM\_TT\_lowTEB\_lensing} (conservative) and \code{base\_mnu\_plikHM\_TT\_lowTEB\_lensing\_BAO} (restrictive).}

The term \likeone{0\nu\beta\beta} is factorized into three terms for GERDA, KamLAND-Zen, and EXO, respectively. Each term is well parametrized by the form
\begin{align}
 \likeone{0\nu\beta\beta,i}=\exp\left[-\left(a_{0,i} + a_{1,i}/T_{1/2} + a_{2,i}/T_{1/2}^2\right)\right] \;,
\end{align}
where $T_{1/2}$ is the neutrinoless double beta decay half-life in units of $\unit{10^{25}}{a}$. The parameters entering this formula have either been determined directly from published data or confirmed by the experimental collaborations.

The neutrino oscillation data is adopted from the global oscillation fit provided by \code{nu-fit.org}~\cite{Esteban:2016qun}. This collaboration does not provide a public likelihood function depending on all oscillation parameters. We thus approximate the likelihood using a spline-fit of the 1-dimensional projections of their $\Delta \chi^2$ fit onto the relevant parameters, shifting the distributions so they have a minimum at $\Delta \chi^2=0$, and taking the likelihood to be a product of the resulting four factors. Each such factor is given by 
\begin{align}
 \exp\left(- \frac{\Delta \chi^2\left(\theta_i\right)}{2}\right) \;,
\end{align}
where $ \chi^2\left(\theta_i\right)$ is the one dimensional projection onto the parameter $\theta_i$. To account for the slight tendency of oscillation data to favor NO, we then added \emph{one} offset $\Delta \chi^2=0.83$~\cite{Esteban:2016qun} to the product value for IO after re-calibrating the projections. This procedure avoids counting the slight inclination towards NO each time it gets projected onto one parameter. The quoted value of $\Delta \chi^2$ corresponds to a Bayes factor of $1.5$ in favor of NO.\footnote{In future analyses, this procedure could be refined by using the full multiparameter likelihoods or by directly fitting the oscillation data.}

%%%%%%%%%%%%%%%%%%%%%%%%%%%%%%%%%%%%%%%%%%%%%%%%%%%%%%%%%%%%%%%%%%%%%%%%%%
\paragraph{Priors}
%%%%%%%%%%%%%%%%%%%%%%%%%%%%%%%%%%%%%%%%%%%%%%%%%%%%%%%%%%%%%%%%%%%%%%%%%%

\begin{figure*}[t]
\begin{subfigure}[c]{0.45\textwidth}
  \includegraphics[width=0.95\textwidth]{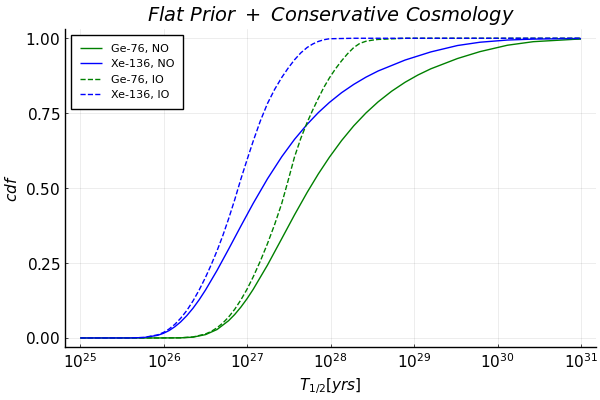}
\end{subfigure}
\begin{subfigure}[c]{0.45\textwidth}
  \includegraphics[width=0.95\textwidth]{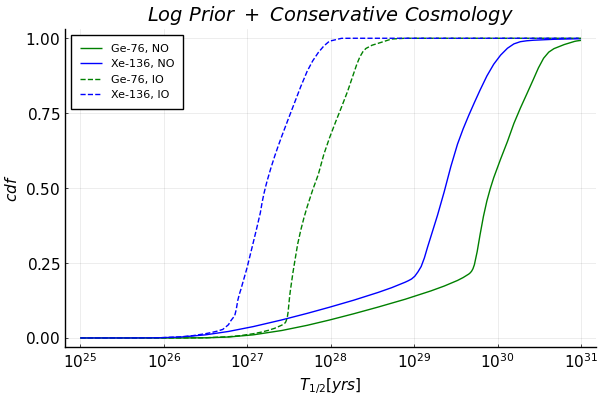}
\end{subfigure}
\caption{\label{fig:T12Posterior}Cumulative distribution function for $T_{1/2}$ for both ${}^{76}$Ge and ${}^{136}$Xe (conservative cosmology). The distributions on the left feature a prior flat in $\sub{m}{lightest}$, while the ones on the right feature a prior flat in $\log(\sub{m}{lightest}$).}
\end{figure*}

Our choice of priors is intended to be informative whenever possible, and otherwise agnostic for parameters with little or no information available. For mass square differences and mixing angles, we choose flat priors extending well beyond the best-fit values for both orderings, since the likelihood functions severely constrain their values in any case. The choice of range does not influence our results. For the Majorana phases $\alpha_1$ and $\alpha_2$, we choose priors flat on the entire range $\left[0, 2\pi \right]$, as we do not have any prior knowledge or theoretical motivation for an informative distribution. The prior on the NME factor $\mathcal{G}$ of a certain isotope is discrete, equally weighing the different theoretical calculations, which deviate from one another by factors of up to three. This procedure will be
further explained and justified in the Appendix (cf. Fig. 4
and subsequent discussion).

%For the critical parameter $\sub{m}{lightest}$, only an upper limit is known. 
We choose two very different priors for $\sub{m}{lightest}$, to explicitly show how results depend on the respective \emph{assumption}.
Both choices span the range $\left[\unit{10^{-7}}{eV}, \unit{0.6}{eV} \right]$.\footnote{The upper end of our range is motivated by the restrictive power of the cosmological data sets: Extending it to larger values does not alter the results significantly. The lower end of the range can be argued to be as low as $\unit{10^{-13}}{eV}$~\cite{Davidson:2006tg}, using a general theoretical argument from perturbation theory. As we are interested in implications for $0\nu\beta\beta$, there will however be a point where lowering the minimal value does not anymore alter any implications.} 
%A flat prior covers the range uniformly, while a scale-free prior covers the range uniform in the logarithm of $\sub{m}{lightest}$.  The following gives a more intuitive description of the different choices of priors: 
The flat prior allocates 90\% of the probability mass at values greater than $\unit{60}{meV}$, while the scale-free log prior allocates about 85\% in a region where $\sub{m}{lightest} < \unit{60}{meV}$.

%%%%%%%%%%%%%%%%%%%%%%%%%%%%%%%%%%%%%%%%%%%%%%%%%%%%%%%%%%%%%%%%%%%%%%%%%%
\section{\label{sec:Results}Results of our analysis}
%%%%%%%%%%%%%%%%%%%%%%%%%%%%%%%%%%%%%%%%%%%%%%%%%%%%%%%%%%%%%%%%%%%%%%%%%%

\begin{figure*}[t]
\begin{subfigure}[c]{0.45\textwidth}
  \includegraphics[width=0.95\textwidth]{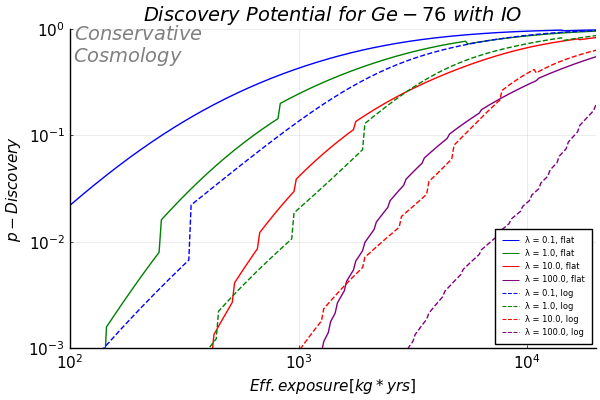}
\end{subfigure}
\begin{subfigure}[c]{0.45\textwidth}
  \includegraphics[width=0.95\textwidth]{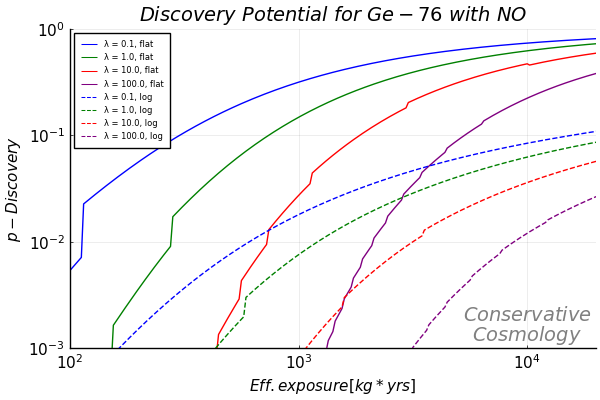}
\end{subfigure}
\\
\begin{subfigure}[c]{0.45\textwidth}
  \includegraphics[width=0.95\textwidth]{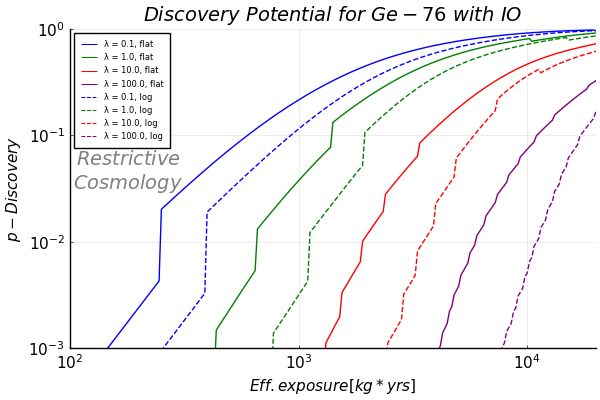}
\end{subfigure}
\begin{subfigure}[c]{0.45\textwidth}
  \includegraphics[width=0.95\textwidth]{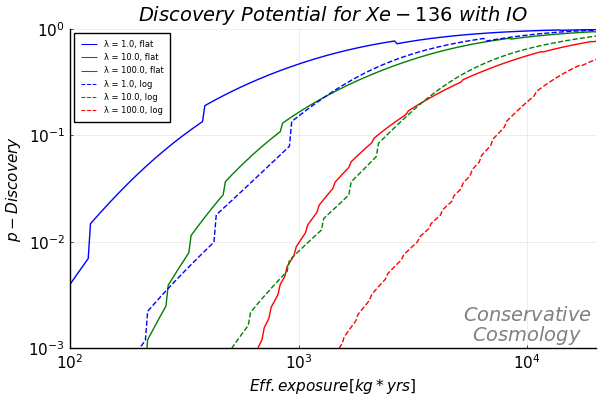}
\end{subfigure}
\caption{\label{fig:DiscoveryPotential}Discovery potential as a function of effective exposure $E \epsilon$ for different background levels. The upper left panel defines a \emph{benchmark case} (germanium, IO, conservative cosmology), while for the other panels one of these parameters is changed: upper right -- NO; lower left -- restrictive cosmology; lower right -- ${}^{136}$Xe. The kinks in the curves are a consequence of the integer nature of Poissonian statistics.}
\end{figure*}

For all combinations of neutrino mass ordering, cosmological data set, and prior for $\sub{m}{lightest}$, we used the BAT Markov chain Monte Carlo code in version \code{v2.0.4}~\cite{Schulz:2021BAT} to sample from the posterior probability density. For each scenario, we ran 4 Markov chains with a length of $1\times10^6$ samples each. The results derived from the posterior samples are presented in \figThreeref{fig:meeSigma}{fig:T12Posterior}{fig:DiscoveryPotential}.

%%%%%%%%%%%%%%%%%%%%%%%%%%%%%%%%%%%%%%%%%%%%%%%%%%%%%%%%%%%%%%%%%%%%%%%%%%
\paragraph{Posterior probabilities}
%%%%%%%%%%%%%%%%%%%%%%%%%%%%%%%%%%%%%%%%%%%%%%%%%%%%%%%%%%%%%%%%%%%%%%%%%%

For fixed mass ordering, cosmology, and prior on $\sub{m}{lightest}$, we show the \emph{posterior probability density} in the plane of the effective mass $|m_{ee}|$ controlling the $0\nu\beta\beta$ rate and the sum $\Sigma$ of neutrino masses in \figref{fig:meeSigma}. We also show theoretical envelopes
bounding the regions in the $\Sigma$--$|m_{ee}|$ parameter space if the Majorana phases are varied within their full ranges while the relevant oscillations parameters are kept at their best fit values (dashed lines) or varied within their $3\sigma$ ranges (solid lines).

Choosing the restrictive cosmological data sets forces $\Sigma$ (and by means of oscillation data also $\sub{m}{lightest}$) to be smaller, resulting in more probability mass shifted towards smaller values of $|m_{ee}|$. Another clear feature is the fact that a scale-free prior on $\sub{m}{lightest}$ leaves almost no probability mass in regions where $\Sigma \gtrsim \unit{0.1}{eV}$. Finally, we note that the highest probability density tends to be at the \emph{upper} ends of the allowed $|m_{ee}|$ ranges for a given $\Sigma$. This reflects the flat prior assumption on the Majorana phases, which makes it highly unlikely to get a cancellation of terms leading to vanishing $|m_{ee}|$~\cite{Lindner:2005kr,Merle:2006du,Maneschg:2008sf}.

%%%%%%%%%%%%%%%%%%%%%%%%%%%%%%%%%%%%%%%%%%%%%%%%%%%%%%%%%%%%%%%%%%%%%%%%%%
\paragraph{Double Beta Decay}
%%%%%%%%%%%%%%%%%%%%%%%%%%%%%%%%%%%%%%%%%%%%%%%%%%%%%%%%%%%%%%%%%%%%%%%%%%

The posterior samples from our global fits can be used to infer the discovery potential for $0\nu\beta\beta$ experiments. The relevant quantity is the half-life probability distribution, examples of which are shown in \figref{fig:T12Posterior}.\footnote{The full set of plots is available in the supplementary material.}  
As expected, the probability mass at smaller half-life is always bigger in the case of IO. However, for the log prior (right panel), favoring small $\sub{m}{lightest}$, the effect is much more pronounced than in the case of the flat prior (left panel).

The relationship between the signal expectation and the half-life is
\begin{equation}
\nu = \frac{N_A \ln 2}{\sub{m}{enr}} \frac{E \epsilon}{T_{1/2}},
\end{equation}
where $N_A$ is Avogadro's number and $\sub{m}{enr}$ is the molar mass of the relevant enriched isotope. $E$ is the exposure and $\epsilon$ is the efficiency to find a signal in the region of interest. The latter includes active/fiducial mass considerations as well as signal reconstruction efficiencies.

Selected results for the discovery potential for $0\nu\beta\beta$, assuming light Majorana neutrinos as the source of a signal, are presented as a function of effective exposure $E \epsilon$ for four different background levels $({\rm BG}=0.1,1,10,100~{\rm events})$ for ${}^{76}$Ge and three for ${}^{136}$Xe $({\rm BG}=1,10,100)$\footnote{Discovery is defined as a Bayes factor of at least $100$ in favor of the presence of $0\nu\beta\beta$. The details of the statistical analysis are available in the supplementary material.} in \figref{fig:DiscoveryPotential}. The discovery potential can be used to judge the merits of individual experiments. In general, we see that a prior flat in $\sub{m}{lightest}$ gives a much larger probability for discovery than a prior flat in $\log(\sub{m}{lightest})$. The importance of keeping the background small is obvious from these plots.

%%%%%%%%%%%%%%%%%%%%%%%%%%%%%%%%%%%%%%%%%%%%%%%%%%%%%%%%%%%%%%%%%%%%%%%%%%
\paragraph{Mass ordering}
%%%%%%%%%%%%%%%%%%%%%%%%%%%%%%%%%%%%%%%%%%%%%%%%%%%%%%%%%%%%%%%%%%%%%%%%%%

As noted earlier, the oscillation data give a Bayes factor of 1.5 in favor of the normal mass ordering.  We have evaluated the Bayes factor that results from including the data of cosmology and $0\nu\beta\beta$ into the analysis for the different cosmological data sets used as well as for our two choices of priors on $\sub{m}{lightest}$. All evidence integrals were
computed using the Cuhre quadrature algorithm provided
by the CUBA library \cite{HAHN200578}.  The results are shown in Table~\ref{tab:BayesFactors}. As can be seen, the additional data and our choices of priors do not change the conclusions significantly. Also the choice
of the cut-off scale at $10^{-7}$ eV is perfectly reasonable, as
lowering it does not change the numerical results significantly anymore.

\begin{table}
 \centering
 \begin{tabular}{|c|c|c|}
  \hline
  Cosmology & Prior on \sub{m}{lightest} & Bayes factor NO/IO \\
  \hline
  restrictive & flat &  1.86 \\
  conservative & flat &  1.64 \\
  restrictive & log &  1.70 \\
  conservative & log &  1.61 \\
  \hline
 \end{tabular}
\caption{ Bayes factors for NO vs IO, for all
combinations of cosmologies and priors.}
\label{tab:BayesFactors}
\end{table}

%%%%%%%%%%%%%%%%%%%%%%%%%%%%%%%%%%%%%%%%%%%%%%%%%%%%%%%%%%%%%%%%%%%%%%%%%%
\section{\label{sec:Conclusions}Conclusions}
%%%%%%%%%%%%%%%%%%%%%%%%%%%%%%%%%%%%%%%%%%%%%%%%%%%%%%%%%%%%%%%%%%%%%%%%%%
In this work, we have conducted a global Bayesian analysis of neutrino mass parameters within the minimal framework of three light Majorana neutrinos, combining data from oscillation experiments, $0\nu\beta\beta$ decay, and precision cosmology. Working with one prior flat in $\sub{m}{lightest}$ and another one flat in $\log\left(\sub{m}{lightest}\right)$, we have investigated both extreme cases of uninformative priors for the critical parameter $\sub{m}{lightest}$. Combining these priors with a more conservative cosmological data set on one hand and a more restrictive one on the other hand, we conclude that the posterior probability for NO is still very mild, even in the  extreme case of a restrictive cosmology and a flat prior on $\sub{m}{lightest}$. In the other cases, the slight inclination towards NO is almost entirely driven by neutrino oscillation experiments.

Furthermore we have evaluated the posterior distributions of $T_{1/2}$ in the different combinations of priors and cosmological data sets as well as for different isotopes. This allowed us to infer the discovery potential of different experimental approaches for $0\nu\beta\beta$. Depending on the neutrino mass ordering, the achievable effective exposure, and the background level, the discovery potential spans a wide range. Assuming a flat prior for $\sub{m}{lightest}$ is in all cases favorable for $0\nu\beta\beta$ searches as is the inverted mass ordering.

Our approach of putting all relevant experimental insights on a consistent and equal footing helps to pave the way for future comprehensive analyses of neutrino data. Additional data can be easily accommodated and probabilities updated in this scheme.

%%%%%%%%%%%%%%%%%%%%%%%%%%%%%%%%%%%%%%%%%%%%%%%%%%%%%%%%%%%%%%%%%%%%%%%%%%
\section*{Acknowledgments}
%%%%%%%%%%%%%%%%%%%%%%%%%%%%%%%%%%%%%%%%%%%%%%%%%%%%%%%%%%%%%%%%%%%%%%%%%%
We would like to thank Matteo Costanzi, Matteo Viel, and in particular Massimiliano Lattanzi for valuable discussions concerning cosmology. Furthermore we are grateful to Itaru Shimizu and Luciano Pandola for checking our likelihood parametrizations for KamLAND-Zen and GERDA, respectively. We thank Frederik Beaujean for advice regarding the implementation of our model in BAT. AM acknowledges partial support by the European Union's Horizon 2020 research and innovation program under the Marie Sklodowska-Curie grant agreements No.~690575 (InvisiblesPlus RISE) and No.~674896 (Elusives ITN), as well as by the Micron Technology Foundation, Inc. MT acknowledges support by the IMPRS-EPP and by the Studienstiftung des deutschen Volkes.

%%%%%%%%%%%%%%%%%%%%%%%%%%%%%%%%%%%%%%%%%%%%%%%%%%%%%%%%%%%%%%%%%%%%%%%%%%%%%%%%%%%%%%%%%%%%%%%%%%%%%%%%%%%%%%%%%%%%%%%%%%%%%%%%%%%%%%%%%%%%%%%%%%%%%%%%%%%%%%%%%%%%%%%%%%%%%%%%%%%%%%%%%%%%%%%%%%%%%%%%%%%%%%%%%%%%%%%%%%%%%%%%%%%%%%%%%%%%%
%%%%%%%%%%%%%%%%%%%%%%%%%%%%%%%%%%%%%%%%%%%%%%%%%%%%%%%%%%%%%%%%%%%%%%%%%%%%%%%%%%%%%%%%%%%%%%%%%%%%%%%%%%%%%%%%%%%%%%%%%%%%%%%%%%%%%%%%%%%%%%%%%%%%%%%%%%%%%%%%%%%%%%%%%%%%%%%%%%%%%%%%%%%%%%%%%%%%%%%%%%%%%%%%%%%%%%%%%%%%%%%%%%%%%%%%%%%%%
%%%%%%%%%%%%%%%%%%%%%%%%%%%%%%%%%%%%%%%%%%%%%%%%%%%%%%%%%%%%%%%%%%%%%%%%%%%%%%%%%%%%%%%%%%%%%%%%%%%%%%%%%%%%%%%%%%%%%%%%%%%%%%%%%%%%%%%%%%%%%%%%%%%%%%%%%%%%%%%%%%%%%%%%%%%%%%%%%%%%%%%%%%%%%%%%%%%%%%%%%%%%%%%%%%%%%%%%%%%%%%%%%%%%%%%%%%%%%
%\begin{widetext}
\onecolumngrid
\vspace*{1cm}
\begin{center}
 \Large{\bf{Supplementary Material}}
\end{center}
This supplementary material describes how we obtained the likelihood functions, it illustrates the procedure we used to determine discovery potentials, and it also presents our main result (namely the posterior probability densities) in an alternative coordinate system, which only uses observables accessible in laboratory experiments.
\vspace*{1cm}

%%%%%%%%%%%%%%%%%%%%%%%%%%%%%%%%%%%%%%%%%%%%%%%%%%%%%%%%%%%%%%%%%%%%%%%%%%
%%%%%%%%%%%%%%%%%%%%%%%%%%%%%%%%%%%%%%%%%%%%%%%%%%%%%%%%%%%%%%%%%%%%%%%%%%
\section*{Statistical Formulation}
%%%%%%%%%%%%%%%%%%%%%%%%%%%%%%%%%%%%%%%%%%%%%%%%%%%%%%%%%%%%%%%%%%%%%%%%%%
The analysis was performed using Bayes' Theorem:
\begin{equation}
	P(H|D,I) = \frac{P(D|H,I)P_0(H|I)}{P(D|I)},
\end{equation}
where $H$ is the hypothesis to which we are assigning a probability, $D$ represents the data being used to update the probability, and $I$ is the additional information and constraints that have been used to define the probabilities.  $P_0(H|I)$ is the prior probability assignment to $H$.  The denominator can be rewritten using the Law of Total Probability as
\begin{equation}
	P(D|I) = \sum_i \int d\boldvec{\theta} P(D|M_i,\boldvec{\theta},I) P(\boldvec{\theta}|M_i,I) P(M_i|I)
\end{equation}
where $M_i$ represent different models allowed under $I$ subject to $\sum_i P(M_i|I) = 1$ and $\boldvec{\theta}$ are the (possibly different) parameters present in the model. $P(\boldvec{\theta}|M_i,I)$ is also normalized.

In our analysis, we assume that there are three light Majorana neutrinos, and have two models representing the two allowed mass ordering schemes: normal (NO: $m_1 < m_2 < m_3$) or inverted (IO: $m_3 < m_1 < m_2$). We assign equal prior probabilities to the two mass orderings: $P(M_1={\rm NO}|I)=P(M_2={\rm IO}|I)=1/2$. The parameters used are \begin{align}
 \boldvec{\theta}=\left(\sub{m}{lightest}, \Delta m^2_\odot, \Delta m^2_A, s^2_{12}, s^2_{13}, \alpha_1, \alpha_2, \mathcal{G}\right) 
 \label{eq:Def:ParamVectorSupp}
\end{align}
where $\sub{m}{lightest}$ is the smallest neutrino mass eigenvalue and $s^2_{ij}$ denotes the sine squared of the mixing angle $\theta_{ij}$. The Majorana phases $\alpha_1$ and $\alpha_2$ take trivial values if neutrinos are of Dirac nature. The nuclear matrix elements (NMEs) required to calculate rates for $0\nu\beta\beta$ are collectively denoted by $\mathcal{G}$. Note that $\boldvec{\theta}$ does \emph{not} contain the Dirac phase $\sub{\delta}{CP}$ or $s^2_{23}$, as they only affect neutrinos oscillations but not $0\nu\beta\beta$ or cosmology. We assign flat probabilities to all parameters except for $\sub{m}{lightest}$ and $\mathcal{G}$. The latter are discussed in more detail below. For the parameters with flat priors, the ranges were defined either to be the maximal range (for the phases) or to cover a sufficiently wide range so as not to affect the results. For example, in the case of the mixing angles and squared mass differences, the parameters are so strongly constrained by the oscillation data that values even remotely close to the edges are already strongly disfavored.

\begin{figure*}[t!]
\vspace{-0.0cm}
\begin{center}   
\begin{tabular}{lr}
\includegraphics[width=6.8cm]{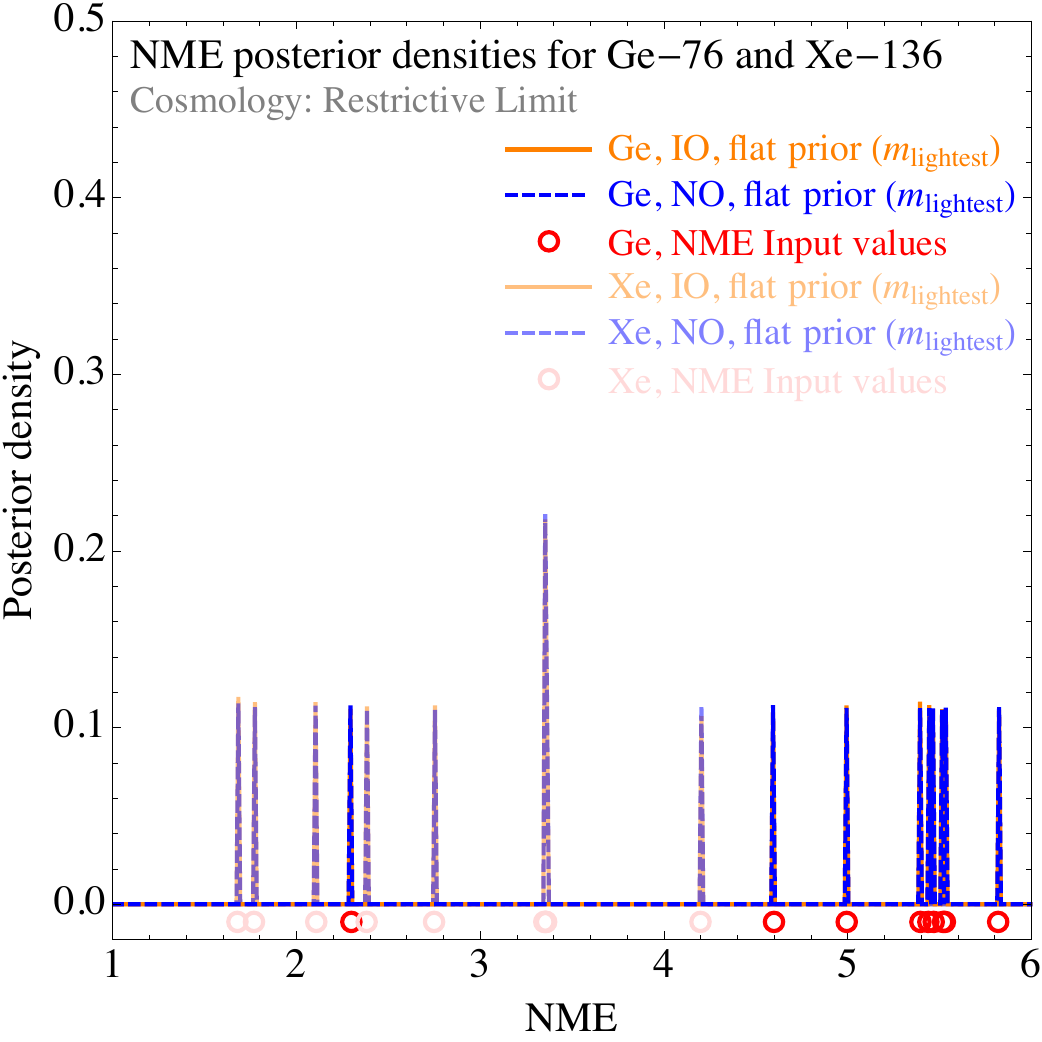} & \includegraphics[width=6.8cm]{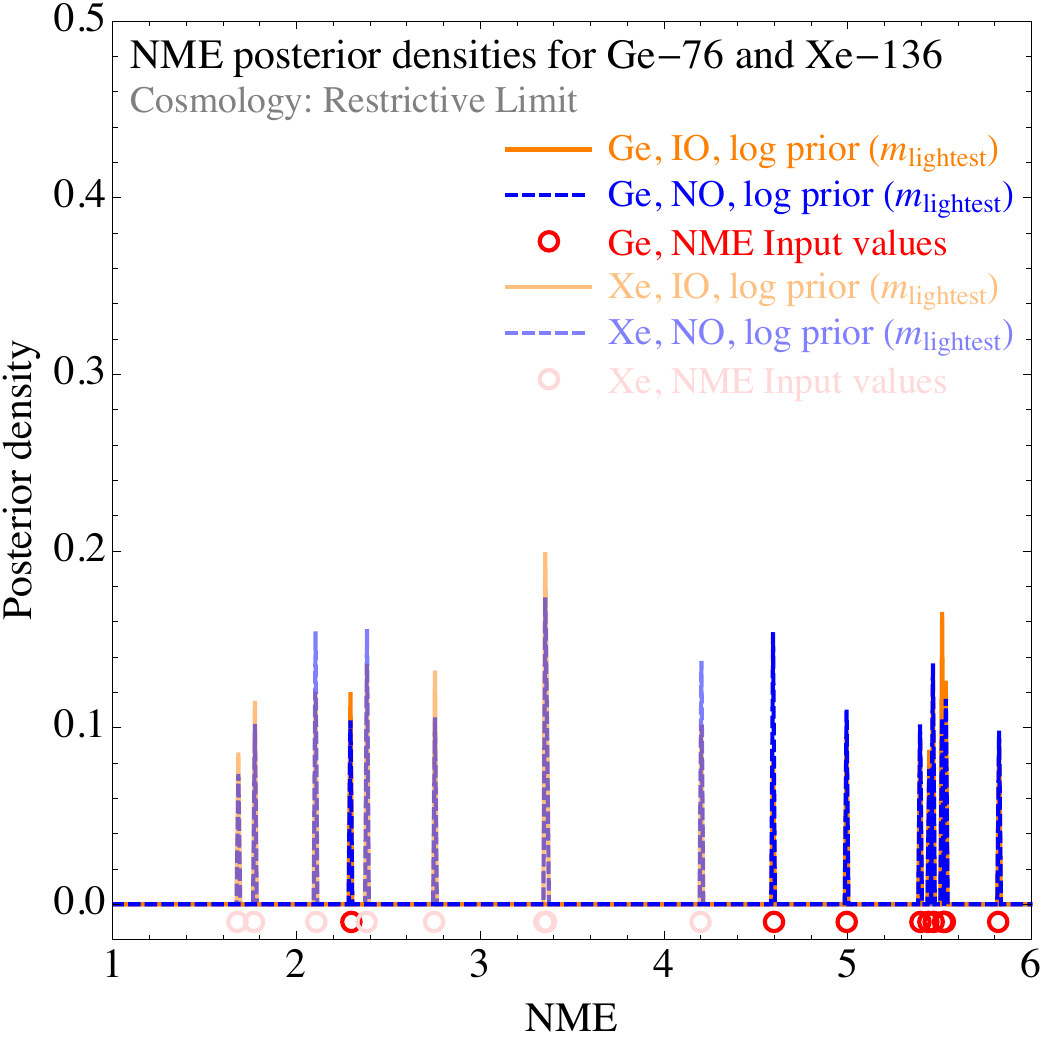}\\
\includegraphics[width=6.8cm]{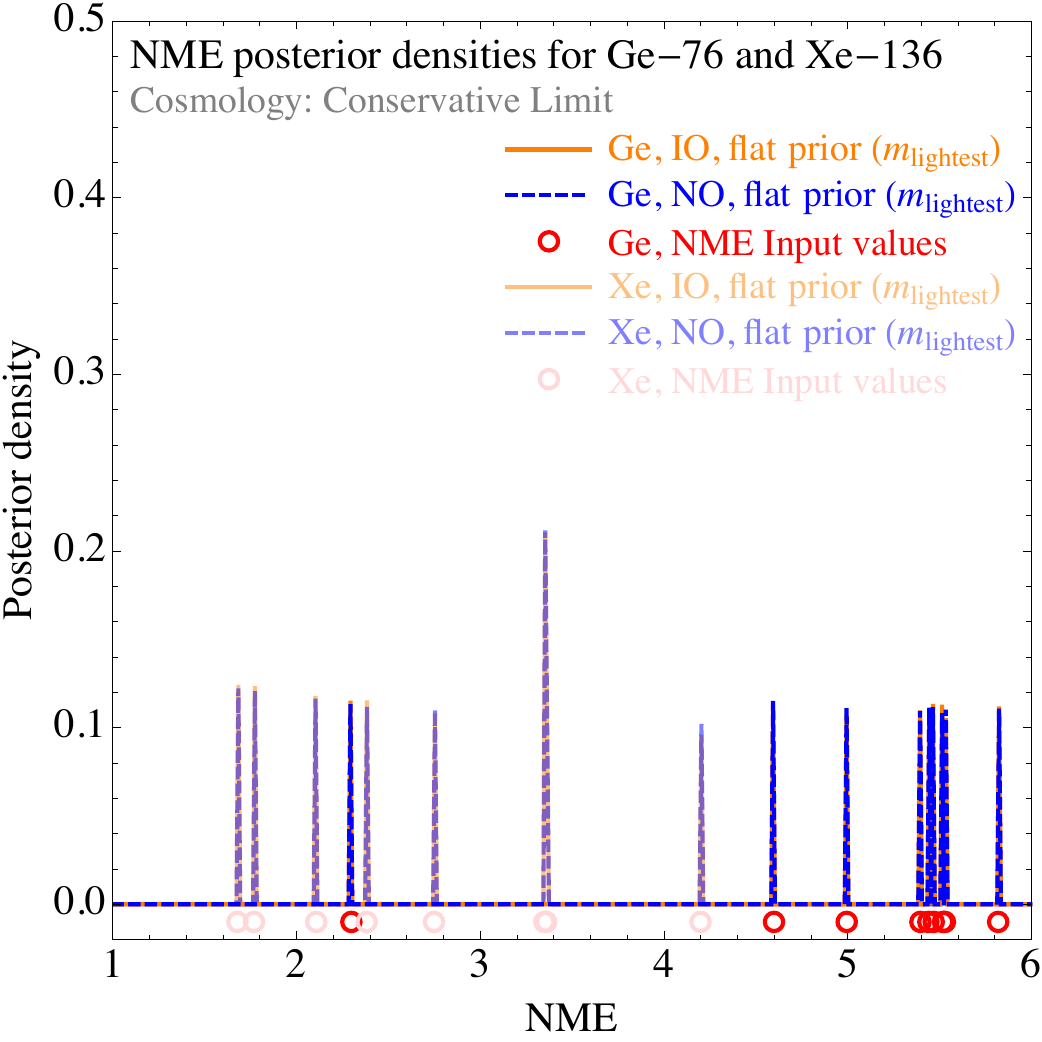} & \includegraphics[width=6.8cm]{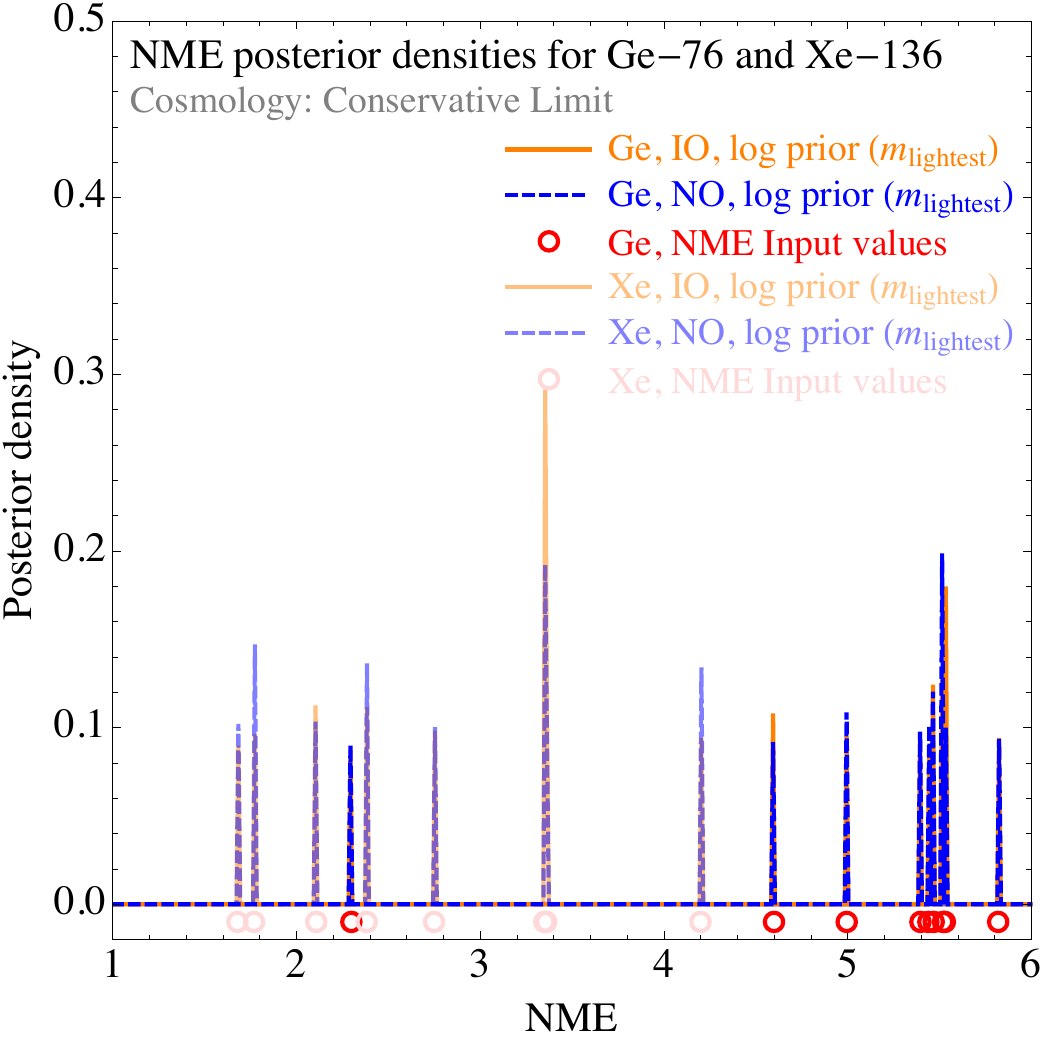}
\end{tabular}
\end{center}
\caption{\label{fig:NME}Posterior probability distributions of the NMEs, sorted into 500 bins, for both ${}^{76}$Ge and ${}^{136}$Xe, along with the input values. As can be seen, the posterior probabilities basically reproduce the input, i.e., the current constraints are not sufficiently strong to impact NME computations.}
\end{figure*}

We consider two prior probability assignments for $\sub{m}{lightest}$: A prior flat in $\sub{m}{lightest}$ (this will tend to favor larger values and it reflects the sizes of the measured mass square differences $\Delta m^2_\odot \equiv m_2^2 - m_1^2$ and $\Delta m^2_A \equiv |m_3^2 - m_1^2|$) and a scale-free prior (flat in the logarithm of $\sub{m}{lightest}$, which tends to favor small values and reflects the fact that the absolute neutrino mass scale is still unknown). We note that
$$P(\log(\sub{m}{lightest})) = {\rm const.} \Leftrightarrow P(\sub{m}{lightest}) \propto 1/\sub{m}{lightest}.$$

The prior probability for $\mathcal{G}$ was defined by assigning equal probabilities of $1/9$ to the different matrix element computations, the results of which are collected in Table~\ref{tab:NMEs}, using the normalization corresponding to the following half-life formula~\cite{Doi:1985dx}:
\begin{equation}
 \frac{1}{T_{1/2}^{0\nu}} = G_{0\nu} |\mathcal{M}_{0\nu}|^2 \left( \frac{|m_{ee}|}{m_e} \right)^2,
 \label{eq:exp_1}
\end{equation}
where $G_{0\nu}$ is a phase space factor which can be easily computed for any isotope under consideration (we make use of the values from Ref.~\cite{Suhonen:1998ck}, which were slightly updated by Ref.~\cite{Rodejohann:2011mu}). $\mathcal{M}_{0\nu}$ is the NME, which encodes all nuclear physics that goes into the process. Some remarks are at order concerning the values shown in Tab.~\ref{tab:NMEs}. For definiteness, we have adopted the standard value $g_A = 1.25$ for the axial vector coupling, which means that we had to rescale some of the NMEs~\cite{Menendez:2008jp,Simkovic:2013qiy}. Also the phase space factors used~\cite{Suhonen:1998ck,Rodejohann:2011mu} correspond to $g_A = 1.25$. In cases where different versions of a computations are available, we have for definiteness always chosen the most optimistic result. For example, Ref.~\cite{Meroni:2012qf} reports the results for both, an intermediate size model space and a large size single particle space, the latter of which tends to yield larger values of the NME. Hence, we have decided to use the large size results. This differs from the choices e.g.\ made in Ref.~\cite{Dev:2013vxa}, where in some cases the smaller and in others the larger value has been chosen. Nevertheless, none of the treatments is wrong in the sense that at the moment the values still suffer from nuclear physics uncertainties, and there exists no way to decide which value is closer to reality.

The NMEs are shown together with their posterior probabilities in Fig.~\ref{fig:NME}. While we only show results for ${}^{76}$Ge and ${}^{136}$Xe, it is straightforward to extend the analysis to any isotope for which NMEs have been calculated.  Note
that the posterior distribution for the NMEs tracks the prior
distribution quite closely. This is to be expected, as the data
from the $0\nu \beta \beta$ experiments does not yet hint towards a
signal but only gives lower bounds on the lifetimes of the
isotopes. Thus, no insights about the NMEs mediating the
decays can be generated, and our prior knowledge does
not get updated in a relevant way. Due to this argument, we
can also justify taking a prior distribution sharply centered
around discrete values. Even if each calculation were
equipped with an uncertainty, which could be accounted
for by modeling the prior by a sum of narrow Gaussians,
this would not change the results significantly. As the
spread between the different calculations is presumably
much larger than this uncertainty, the overall picture would
stay unaltered. The choice of assigning a equal normalization of $N^{-1}_{calc}$ to each calculation just formalizes our
current state of ignorance concerning the best method—a
state that can only be changed by observation of
$0\nu \beta \beta$ decay.

\begin{figure*}[bh]
\begin{center}   
\begin{tabular}{lr}
\includegraphics[width=7.2cm]{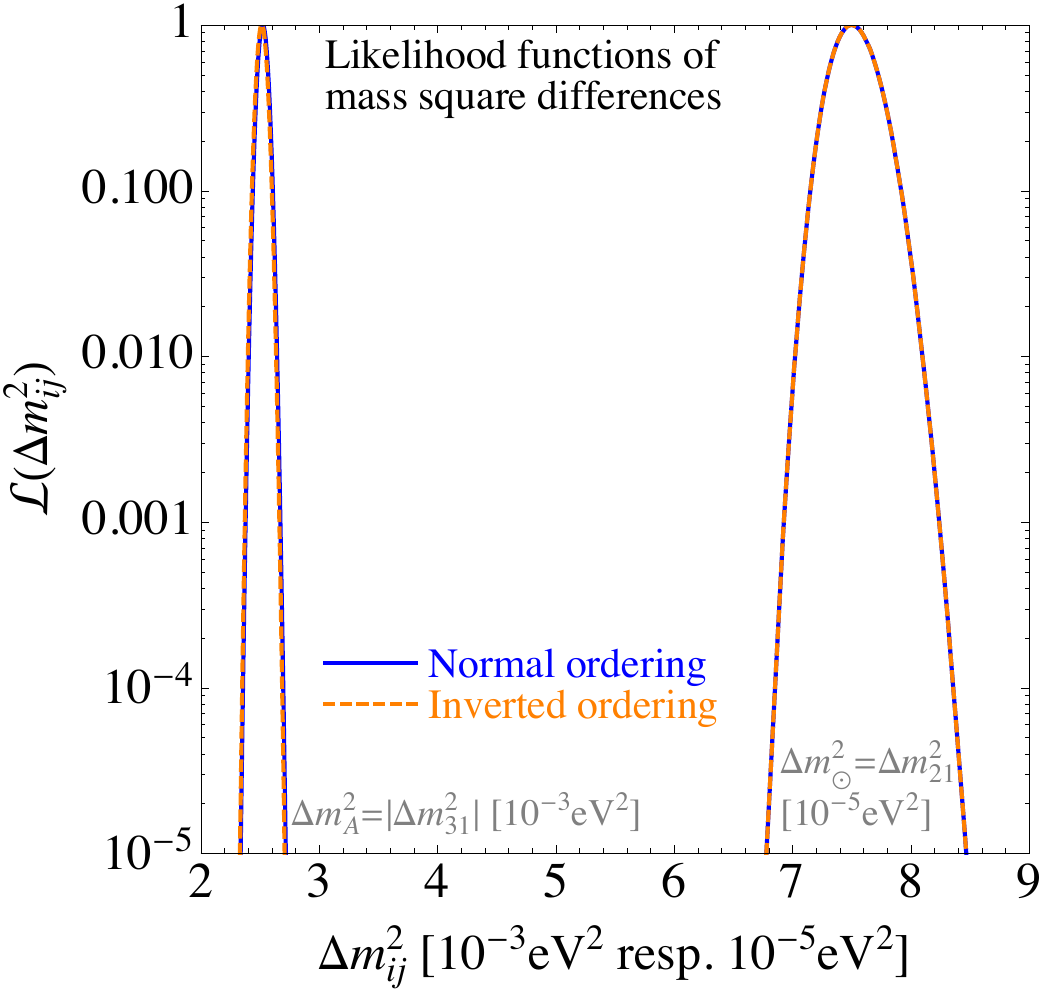} & \includegraphics[width=7.2cm]{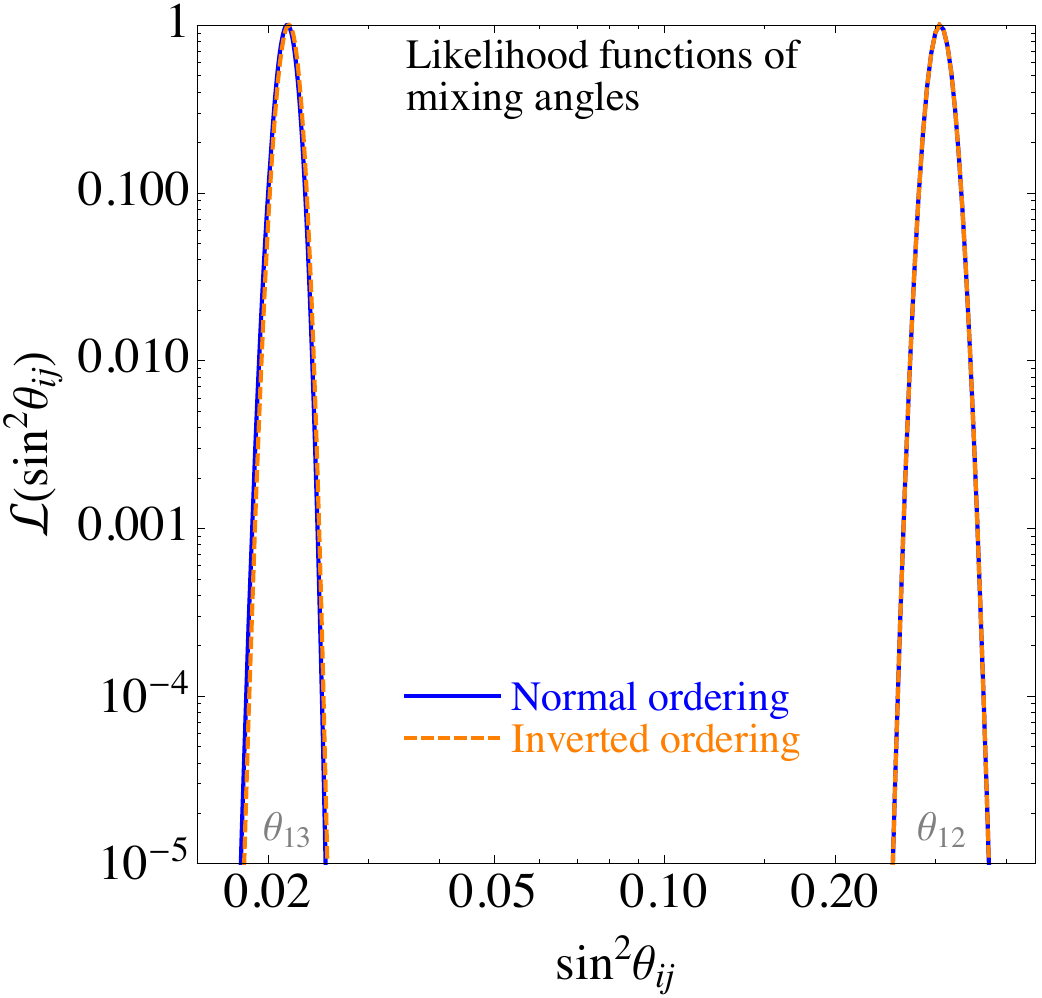}
\end{tabular}
\end{center}
\caption{\label{fig:osc-like}Likelihood functions for the squared mass differences (left) and for the relevant mixing angles (right).}
\end{figure*}

\begin{table*}[bht]
\hspace{-1cm}
\begin{tabular}{|c||c|c|}\hline
Method [Reference] & Value for ${}^{76}$Ge & Value for ${}^{136}$Xe\\\hline\hline
Energy Density Functional Method~\cite{Rodriguez:2010mn}
& $4.60$ & $4.20$ \\
Interacting Shell Model~\cite{Menendez:2008jp}
& $2.30$ & $1.77$ \\
Microscopic Interacting Boson Model~\cite{Barea:2013bz}
& $5.47$ & $3.36$ \\
Proton-Neutron QRPA~\cite{Suhonen:2010zzc}
& $5.52$ & $3.35$ \\
Self-Consistent Renormalized QRPA (Bonn potential)~\cite{Meroni:2012qf}
& $5.82$ & $3.36$ \\
Self-Consistent Renormalized QRPA (Argonne V18 potential)~\cite{Meroni:2012qf}
& $5.44$ & $2.75$ \\
QRPA (Bonn potential)~\cite{Simkovic:2013qiy}
& $5.40$ & $2.38$ \\
QRPA (Argonne V18 potential)~\cite{Simkovic:2013qiy}
& $5.00$ & $2.11$ \\
Deformed Self-Consistent Skyrme QRPA~\cite{Mustonen:2013zu}
& $5.53$ & $1.68$\\ \hline
\end{tabular}
\caption{\label{tab:NMEs}NMEs as extracted from the references given. See text for details.}
\end{table*}

The probability of the data, $P(D|M_i,\boldvec{\theta},I)$, viewed as function of the parameters with the data fixed, is known as the likelihood $\mathcal{L}(\boldvec{\theta})$. In the following, we describe which prescriptions were used to define these likelihoods.

%%%%%%%%%%%%%%%%%%%%%%%%%%%%%%%%%%%%%%%%%%%%%%%%%%%%%%%%%%%%%%%%%%%%%%%%%%
\subsection*{Neutrino Oscillation Data}

The neutrino oscillation likelihood is taken from {\tt nu-fit.org}~\cite{Esteban:2016qun}, of which we have used {\tt v3.0} of their neutrino oscillation global fit results. As described in the main text, we have converted the $\Delta \chi^2$ projections from {\tt nu-fit.org} into likelihood functions, $\mathcal{L}(\theta_i) \propto \exp \left[ -\Delta \chi^2(\theta_i)/2 \right]$, where $\Delta \chi^2(\theta_i)$ is the $\Delta \chi^2$ projection for the parameter $\theta_i$. The likelihoods are shown in Fig.~\ref{fig:osc-like}. Note that the fit by {\tt nu-fit.org} very slightly disfavors IO, by an offset $\Delta \chi^2 = 0.83$. However, while this value appears in each of the projections shown by that collaboration, we have to be careful not to penalize IO more than once in our analysis, because the value of $0.83$ also appears only once in the full $\chi^2$-function, before projecting out individual parameters. We have accounted for this by adding the offset only \emph{once}, to avoid an artificial worsening of IO which would even increase with the number of parameters used. Note that this results in a global factor in front of the likelihood function for IO, which drops out when computing posterior probabilities. However, once we compute a Bayes Factor, both likelihoods need to share a common normalization to make sense of the values obtained, and in that case this constant prefactor does play a role, and it corresponds to the factor of $1.5$ in favor of NO, as mentioned in the main text.

\begin{figure*}[th]
\begin{center}   
\begin{tabular}{lr}
\includegraphics[width=7.4cm]{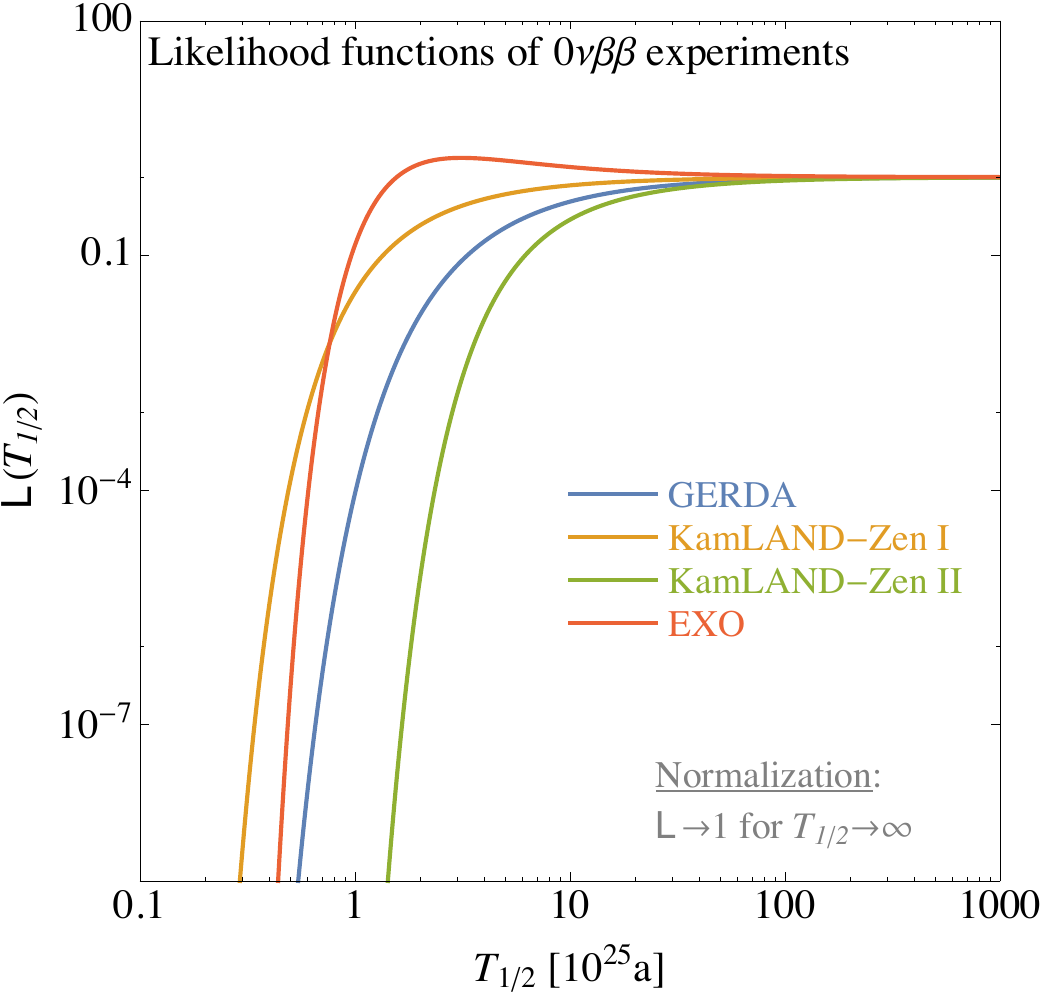} & \includegraphics[width=7.0cm]{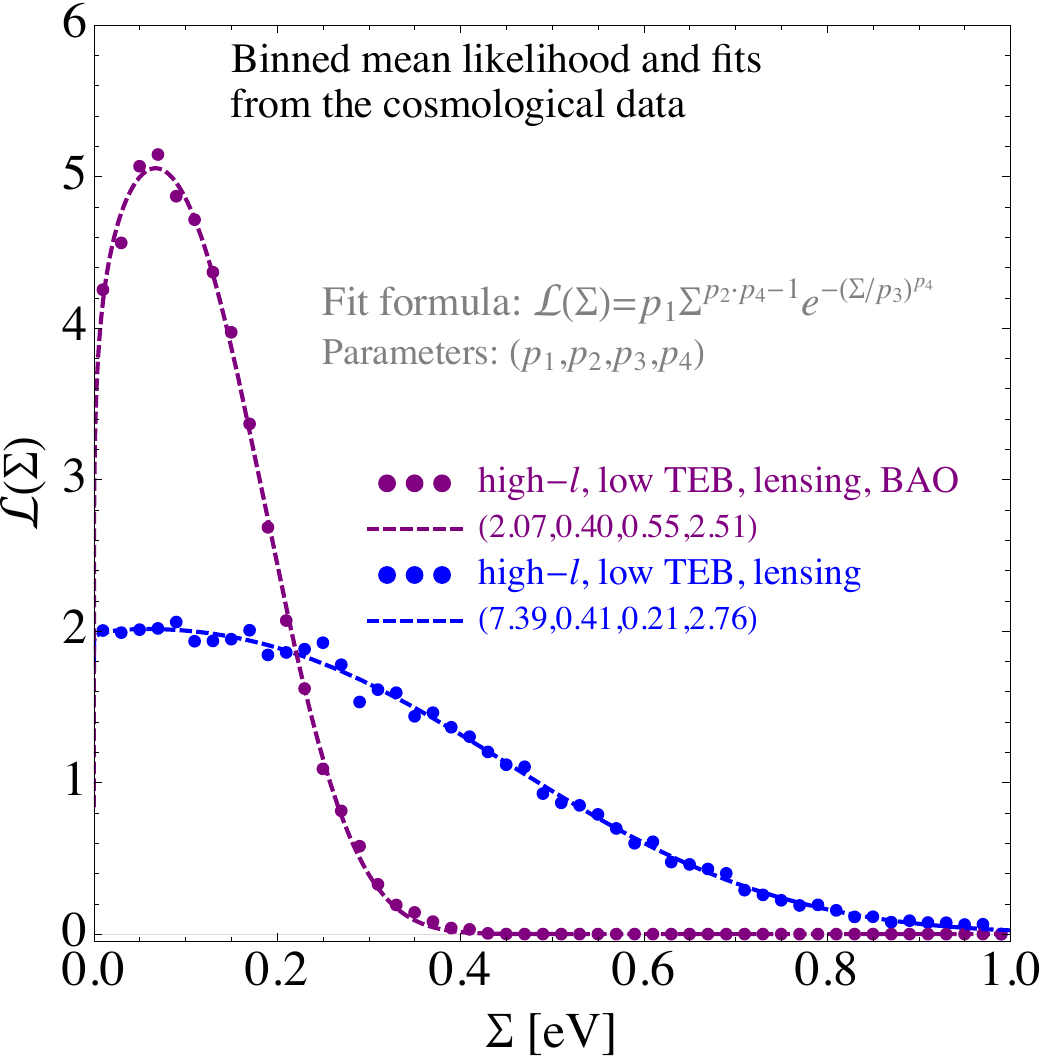}
\end{tabular}
\end{center}
\caption{\label{fig:like_0nbbCosmo}Likelihood functions for the different $0\nu\beta\beta$ experiments (left) and cosmological data sets (right).}
\end{figure*}

%%%%%%%%%%%%%%%%%%%%%%%%%%%%%%%%%%%%%%%%%%%%%%%%%%%%%%%%%%%%%%%%%%%%%%%%%%
\subsection*{Double Beta Decay data}

The most strongly constraining experiments to date are the GERDA, KamLAND-Zen, and EXO experiments, and we have included likelihood terms for these experiments in our analysis. The data were analyzed and presented in different ways by the three experiments. We discuss these in turn to explain how we included these data into our Bayesian analysis. Note that $T_{1/2}$ values are always given in units of $10^{25}$~years. The functions we will present in the following are plotted in the left panel of Fig.~\ref{fig:like_0nbbCosmo} for illustration.

%%%%%%%%%%%%%%%%%%%%%%%%%%%%%%%%%%%%%%%%%%%%%%%%%%%%%%%%%%%%%%%%%%%%%%%%%%
\subsubsection*{GERDA}

The GERDA Collaboration~\cite{ref:GERDA} performed both a Bayesian and a frequentist limit setting analysis on their data.  We use the posterior probability density for the inverse of the half-life of $^{76}$Ge as determined by the GERDA Collaboration, which is well parametrized by the form\footnote{The parametrizations have been cross-checked by the collaboration (L.~Pandola, private communication).}
\begin{equation}
\mathcal{P}(1/T^{\rm Ge}_{1/2}) \propto \exp\left[-\frac{1}{2} \frac{(1/T^{\rm Ge}_{1/2}+1.48)^2}{0.461^2}\right] \; .
  \end{equation}
Since GERDA used a flat prior for $1/T^{\rm Ge}_{1/2}$, we use
\begin{equation}
\mathcal{L}_{\rm GERDA}(T^{\rm Ge}_{1/2}) \propto \exp\left[-\frac{1}{2} \frac{(1/T^{\rm Ge}_{1/2}+1.48)^2}{0.461^2}\right].
\end{equation}

%%%%%%%%%%%%%%%%%%%%%%%%%%%%%%%%%%%%%%%%%%%%%%%%%%%%%%%%%%%%%%%%%%%%%%%%%%
\subsubsection*{KamLAND-Zen}

The KamLAND-Zen collaboration~\cite{ref:Kamland-Zen} performed frequentist analyses based on a profile likelihood test statistic. They presented their results in the form of $\Delta \chi^2$ as a function of $T^{\rm Xe}_{1/2}$ (Wilk's Theorem was assumed to apply). We have interpreted these $\Delta \chi^2$ as likelihoods for the half-life of $^{136}$Xe by setting
\begin{equation}
\mathcal{L}_\text{KamLAND-Zen}(T^{\rm Xe}_{1/2}) \propto \exp\left[-\frac{1}{2} \Delta \chi^2 (T^{\rm Xe}_{1/2}) \right] \; .
  \end{equation}  
 The Phase-I and Phase-II results were used separately, and the following parametrizations were found to represent the KamLAND-Zen results well:\footnote{The parametrizations have been cross-checked by the collaboration (I.~Shimizu, private communication).}
 \begin{equation}
\Delta \chi^2 =
\left\{
\begin{matrix}
2 \left[2.3/T^{\rm Xe}_{1/2} + 1.09/(T^{\rm Xe}_{1/2})^2 \right] \;\;\;\; {\rm Phase-I,} \hfill\hfill\hfill\hfill\hfill \\[2mm]
2 \left[9.71/T^{\rm Xe}_{1/2} + 28.1/(T^{\rm Xe}_{1/2})^2 \right] \;\; {\rm Phase-II.} \hfill\hfill\hfill\hfill\hfill
\end{matrix}\right.
\end{equation}

%%%%%%%%%%%%%%%%%%%%%%%%%%%%%%%%%%%%%%%%%%%%%%%%%%%%%%%%%%%%%%%%%%%%%%%%%%  
\subsubsection*{EXO}

The EXO Collaboration also performed a frequentist analysis and presented confidence level limits on $T^{\rm Xe}_{1/2}$.  However, the background expectation, background uncertainty, and observed numbers of events were presented in the paper~\cite{ref:EXO}, allowing for the construction of the likelihood as
\begin{equation}
\mathcal{L}_{\rm EXO}(T^{\rm Xe}_{1/2}) \propto \int e^{-(\nu+\lambda)}\frac{(\nu+\lambda)^n}{n!} \mathcal{G}(\lambda; \lambda_0=31.1,\sigma_\lambda=3.3) d\lambda,
\end{equation} 
where the signal expectation $\nu$ depends on $T^{\rm Xe}_{1/2}$ as well as on the exposure and the efficiency, which are both given in Ref.~\cite{ref:EXO}. The background level is given by $\lambda$ and is centered on $\lambda_0$. The uncertainty is introduced by smearing the expectation with a Gaussian of width $\sigma_\lambda$. The observed number of events is $n=39$.  The resulting likelihood is well parametrized by the form
 \begin{equation}
\mathcal{L}_{\rm EXO}(T^{\rm Xe}_{1/2}) \propto \exp\left[-\frac{1}{2} \frac{(1/T^{\rm Xe}_{1/2}-0.32)^2}{0.30^2}\right] \;.
  \end{equation}
We note that the peak is at finite $T^{\rm Xe}_{1/2}$ since more events were observed than expected from background processes.

%%%%%%%%%%%%%%%%%%%%%%%%%%%%%%%%%%%%%%%%%%%%%%%%%%%%%%%%%%%%%%%%%%%%%%%%%%
\subsection*{Cosmological Data sets}
%%%%%%%%%%%%%%%%%%%%%%%%%%%%%%%%%%%%%%%%%%%%%%%%%%%%%%%%%%%%%%%%%%%%%%%%%%
%%%%%%%%%%%%%%%%%%%%%%%%%%%%%%%%%%%%%%%%%%%%%%%%%%%%%%%%%%%%%%%%%%%%%%%%%%

The cosmological factor $\mathcal{L}_{\rm cosmo}$ contains extensive information about cosmic structure formation and cannot be given in a simple closed form. Instead, we have used Markov chains sampling the posterior for different combinations of datasets that are publicly available on the Planck Legacy Archive (PLA).\footnote{Based on observations by Planck (\url{http://www.esa.int/Planck}).} These posterior samples were obtained from a uniform prior in $\Sigma$, such that the posterior is identical to the likelihood except for some unknown normalization that is irrelevant in our implementation. The total number of samples for each model is about $50\times 10^3$. The samples were binned into 50 bins to obtain a likelihood estimator, which was then fit to the following four-parameter template:
\begin{align}
        \like{cosmo}{\Sigma} = p_1 \, \Sigma^{p_2 p_3 -1} \, \exp\left[-\left(
\frac{\Sigma}{p_3}\right)^{p_4}\right] \;.
        \label{eq:LikeCosmoParam}        
\end{align}   
This functional form is related to the generalized Gamma distribution and it is a good empirical fit to likelihood, as the right panel of Fig.~\ref{fig:like_0nbbCosmo} shows. \Tabref{tab:CosmoLikeParamValues} lists the fit coefficients we obtained for
the two cosmological data sets used in our analysis.

\begin{table*}[th]
        \centering
        \begin{tabular}{|c||c|c|c|c|c|}
        \hline
        Level & Data set & $p_1$ & $p_2$ & $p_3$ & $p_4$ \\
        \hline \hline
        Conservative & \code{base\_mnu\_plikHM\_TT\_lowTEB\_lensing} &  $2.073$  &  $0.401$  &  $0.552$  &
$2.514$ \\
        \hline
        Restrictive & \code{base\_mnu\_plikHM\_TT\_lowTEB\_lensing\_BAO} &  $7.385$  &  $0.407$  & 
$0.207$  & $2.765$ \\
        \hline        
        \end{tabular}
        \caption{\label{tab:CosmoLikeParamValues}Fit parameters for the cosmological likelihood factor parametrized by Eq.~\eqref{eq:LikeCosmoParam} for both data sets used in our analysis.}
\end{table*}

%%%%%%%%%%%%%%%%%%%%%%%%%%%%%%%%%%%%%%%%%%%%%%%%%%%%%%%%%%%%%%%%%%%%%%%%%%
%%%%%%%%%%%%%%%%%%%%%%%%%%%%%%%%%%%%%%%%%%%%%%%%%%%%%%%%%%%%%%%%%%%%%%%%%%
\section*{Double Beta Decay: Discovery Probability Analysis}
%%%%%%%%%%%%%%%%%%%%%%%%%%%%%%%%%%%%%%%%%%%%%%%%%%%%%%%%%%%%%%%%%%%%%%%%%%

We define two hypotheses:

\begin{itemize}
\item[$H_0$] There are only background processes producing the data;
\item[$H_1$] There is, additionally to the background processes, also a signal from neutrinoless double beta decay.
\end{itemize}
The full information is contained in the posterior probability for the hypothesis given the data, $P(H_1|D)$.  If this probability is large enough, then a `discovery' can be claimed.  We will use the related concept of posterior odds
$$\mathcal{O}_1 = \frac{P(H_1|D)}{P(H_0|D)}$$
and place requirements on this quantity to claim a discovery.  Assuming the two hypotheses are exhaustive\footnote{We exclude here the possibility that there are background sources not included in the background model expectations.  We also assume that the signal would come from the exchange of a light Majorana neutrino.} we have
\begin{eqnarray}
P(H_0|D) &=& \frac{P(D|H_0)P_0(H_0)}{P(D|H_1)P_0(H_1)+P(D|H_0)P_0(H_0)},\nonumber \\
P(H_1|D) &=& \frac{P(D|H_1)P_0(H_1)}{P(D|H_1)P_0(H_1)+P(D|H_0)P_0(H_0)}.\nonumber
\end{eqnarray}
The posterior odds are then given by
$$\mathcal{O}_1 = \frac{P(D|H_1)}{P(D|H_0)}\mathcal{O}_0$$
where 
$$\mathcal{O}_0 = \frac{P_0(H_1)}{P_0(H_0)}$$
are the prior odds.  The factor
$\frac{P(D|H_1)}{P(D|H_0)}$ is known as the `Bayes Factor'.

We consider the case of a single experiment measuring $n$ events in a given region of interest defined by the $Q$ value of the decay, the background level, and the energy resolution.  We use simple Poisson statistics to evaluate the discovery potential.  Small improvements are possible by making use of the background and signal spectral shapes, but the additional sensitivity gained by doing this is not significant for our discussion.  We will also ignore uncertainties in the background and assume it is well known. The uncertainty in the background typically has little effect on the discovery calculation if the background is small, but it can become important if this is not the case.

We use the symbol $\lambda$ to represent the background expectation in the region of interest, and $\nu$ is the signal expectation given a half-life for the decay, the exposure, and the detection efficiency.

Given these definitions, we have
\begin{equation}
P(D|H_1) = P(n|H_1) = \int_0^{\infty} P(n|\lambda+\nu')P(\nu') d\nu' \; ,
\end{equation}
where $P(\nu')$ is the probability to have the signal expectation $\nu'$ given the experimental conditions and prior knowledge on $T_{1/2}$. The probability of the data for $H_0$ is simply
\begin{equation}
P(D|H_0) = P(n|H_0) = e^{-\lambda} \frac{\lambda^n}{n!} \; .
\end{equation}

The relationship between the signal expectation and the half-life is
\begin{equation}
\nu = \frac{N_A \ln 2}{m_{\rm enr}} \frac{E \epsilon}{T_{1/2}},
\end{equation}
where $N_A$ is Avogadro's number and $m_{\rm enr}$ is the molar mass of the relevant enriched isotope.  $E$ is the exposure and $\epsilon$ is the efficiency to find a signal in the region of interest. The latter includes active/fiducial mass considerations as well as signal reconstruction efficiencies.

\begin{figure*}[th]
\vspace{-0.0cm}
    \centering
    \begin{tabular}{lr}
\includegraphics[width=7.2cm]{figures/Thalflife_flat_conservative.png} & \includegraphics[width=7.2cm]{figures/Thalflife_log_conservative.png}\\
\includegraphics[width=7.2cm]{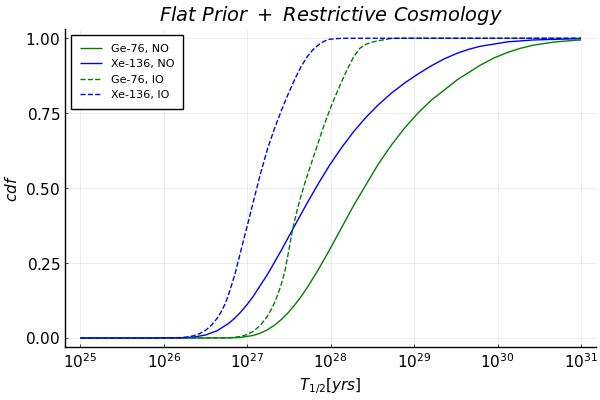} & \includegraphics[width=7.2cm]{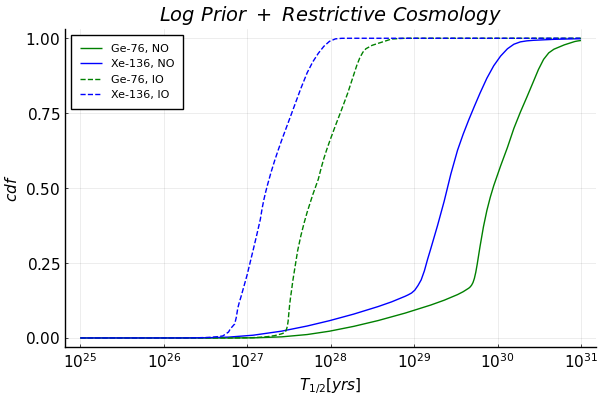}
\end{tabular}
    \caption{\label{fig:PT}Cumulative probability distributions for $T_{1/2}$ for both ${}^{76}$Ge and ${}^{136}$Xe, featuring both mass orderings and both priors on the lightest neutrino mass. As to be expected, the logarithmic prior disfavors NO more strongly than the flat prior.}
 \end{figure*}

We evaluate the probability of a claim of discovery assuming three light Majorana neutrinos by:
\begin{eqnarray}
\label{eq:discover}
P_{\rm discovery} & =& \sum_{n=0}^{\infty}    P(n|H_1)  I(\mathcal{O}_1),\nonumber \\
P_{\rm discovery} & =& \sum_{n=0}^{\infty} \left[ \int_0^{\infty}  P(n|\nu'+\lambda) P(\nu') d\nu' \right] I(\mathcal{O}_1),\nonumber
\end{eqnarray}
where $I$ is the indicator function: 
\begin{eqnarray*}
I&=&1 \;\;\;\;\; {\rm if} \;\;\;\;\;   \mathcal{O}_1\geq 100 \; , \\
I&=&0 \;\;\;\;\; {\rm if} \;\;\;\;\;   \mathcal{O}_1< 100 \; .
\end{eqnarray*}
In addition to the exposure and the signal efficiency, the discovery potential depends on the background level and on the cut value for the posterior odds. We take the prior odds to be $1$. The probability distribution for $\nu$ is given by the results of our neutrino parameter analysis, and depends on the mass orderings, on the probability distribution for the effective neutrino mass in these orderings, and on the matrix element values.  We give the probability distributions for $T_{1/2}$ for $^{76}$Ge and $^{136}$Xe for the different orderings, cosmological data sets sets, and mass scale priors in Fig.~\ref{fig:PT}.

 \begin{figure*}[th]
\centering
\begin{tabular}{lr}
\includegraphics[width=7.2cm]{figures/Discovery_Conservative_Ge_IO_1.png} & \includegraphics[width=7.2cm]{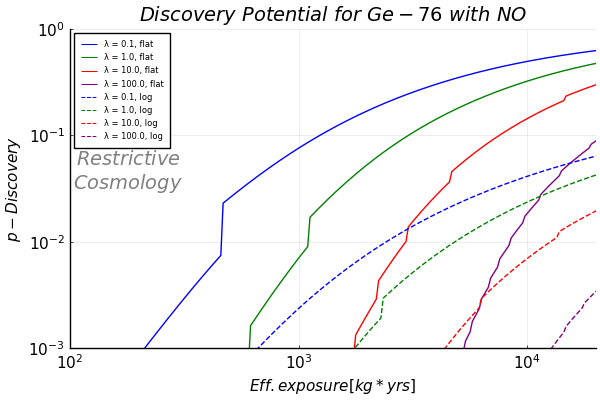}\\
\includegraphics[width=7.2cm]{figures/Discovery_Conservative_Ge_IO_2.png} & \includegraphics[width=7.2cm]{figures/Discovery_Conservative_Ge_NO_2.png}
\end{tabular}
\caption{\label{fig:discover_1}Discovery potential for ${}^{76}$Ge: IO/NO (left/right) and restrictive/conservative cosmology (top/bottom).}
\end{figure*}

\begin{figure*}[th]
\centering
\begin{tabular}{lr}
\includegraphics[width=7.2cm]{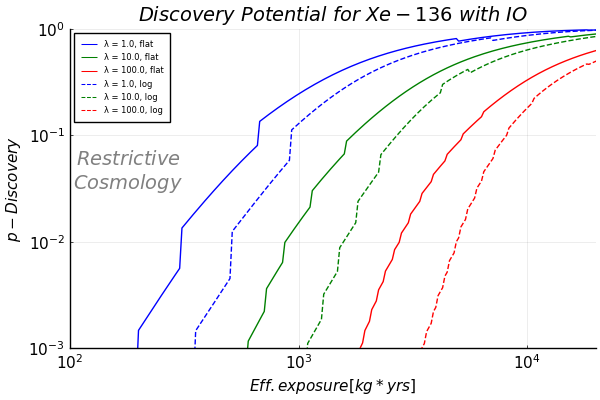} & \includegraphics[width=7.2cm]{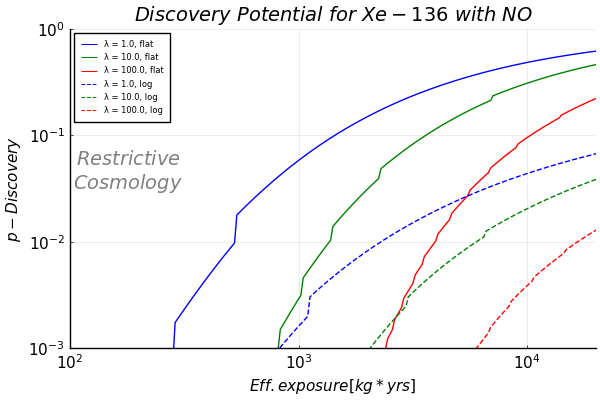}\\
\includegraphics[width=7.2cm]{figures/Discovery_Conservative_Xe_IO_2.png} & \includegraphics[width=7.2cm]{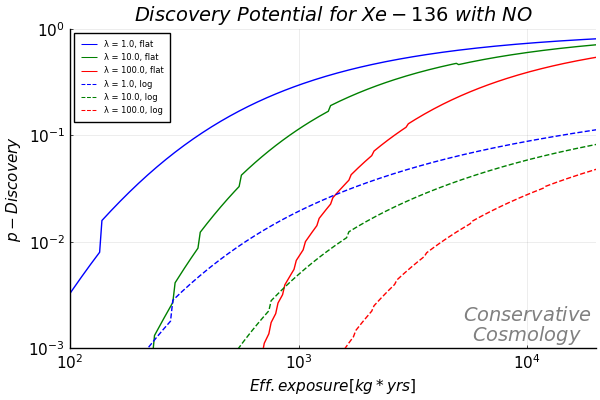}
\end{tabular}
\caption{\label{fig:discover_2}Discovery potential for ${}^{136}$Xe: IO/NO (left/right) and restrictive/conservative cosmology (top/bottom).}
\end{figure*}
 
We use the output of BAT to carry out the integrals:
\begin{eqnarray}
P(n|H_1) &=& \int_0^{\infty} P(n|\lambda+\nu')P(\nu') d\nu' \;  \\
&=& E [ P(n|\lambda+\nu) ]_{P(\nu)} \\
&\approx& \frac{1}{N} \sum_{i=1}^{N} P(n|\lambda+\nu)|_{P(\nu)},
\end{eqnarray}
with $N$ being the number of samples of $T_{1/2}$ taken from the Markov chain. Recall that the background level $\lambda$ is fixed.  

The discovery potential is presented as a function of efficiency reduced exposure, $E \epsilon$, for four different background levels $(\lambda=0.1,1,10,100)$ for Ge and three for Xe $(1,10,100)$. The results are shown in Figs.~\ref{fig:discover_1} and~\ref{fig:discover_2}.\footnote{Results for other isotopes can be requested from the authors.}

The cut value for the Bayes Factor was taken to be $100$ -- i.e., we would claim a discovery if $H_1$ was 100 times more probable than $H_0$. This is an arbitrary choice, and varying it by a factor $10$ is certainly possible, but it does not have strong influence on the results of this analysis. The jumps in the curve are due to the Poisson nature of the statistical fluctuations.

%%%%%%%%%%%%%%%%%%%%%%%%%%%%%%%%%%%%%%%%%%%%%%%%%%%%%%%%%%%%%%%%%%%%%%%%%%
\section*{Posterior probability for lab-based experiments}
%%%%%%%%%%%%%%%%%%%%%%%%%%%%%%%%%%%%%%%%%%%%%%%%%%%%%%%%%%%%%%%%%%%%%%%%%%

In the main text, we have shown the heat map of posterior probability density in the $\Sigma$--$|m_{ee}|$ plane, i.e., in terms of the observables that are probed by cosmology and by $0\nu\beta\beta$. This choice is inspired by these two observables being the most promising both to identify the neutrino mass ordering and to get information on the absolute neutrino mass scale. In the theoretical literature, one often finds plots of $m_{\rm lightest}$ vs.\ $|m_{ee}|$, see e.g.\ the literature on how to probe certain neutrino mass models by $0\nu\beta\beta$~\cite{King_2013,PhysRevD.94.093003,Gehrlein_2015,Agostini_2016}. However, this choice is problematic in our case because not only is $m_{\rm lightest}$ only a theoretical parameter rather than an actual observable, but it also exhibits a very strong dependence on the prior, which is poorly constrained.

\begin{figure*}[h!]
\vspace{-0.0cm}
\centering
\begin{tabular}{lr}
\includegraphics[width=7.2cm]{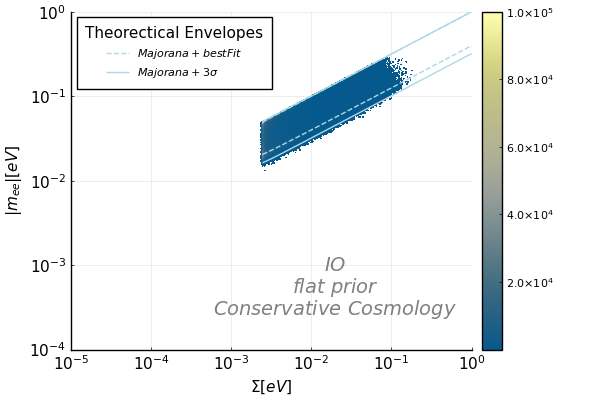} & \includegraphics[width=7.2cm]{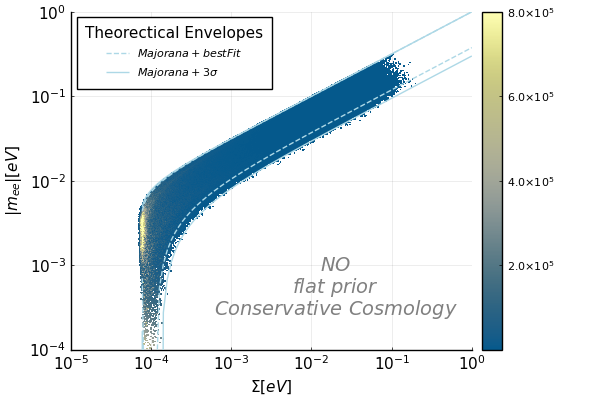}
\\
\includegraphics[width=7.2cm]{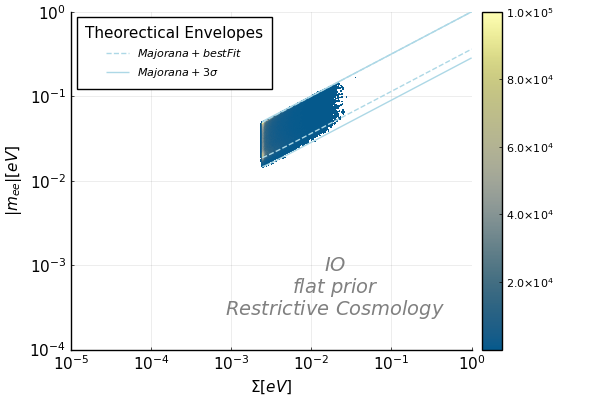} & \includegraphics[width=7.2cm]{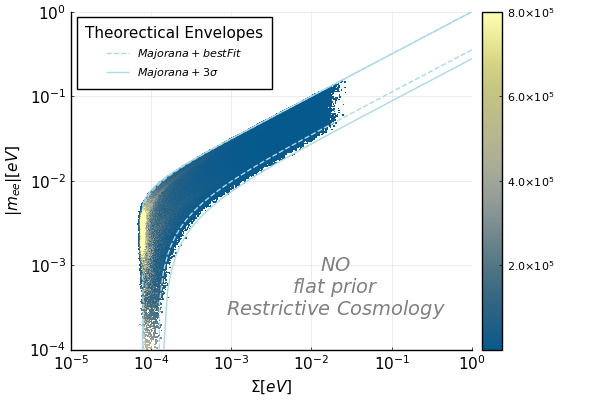}
\\
\includegraphics[width=7.2cm]{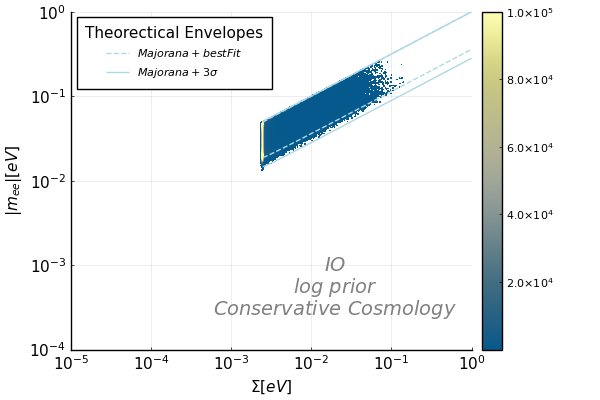} & \includegraphics[width=7.2cm]{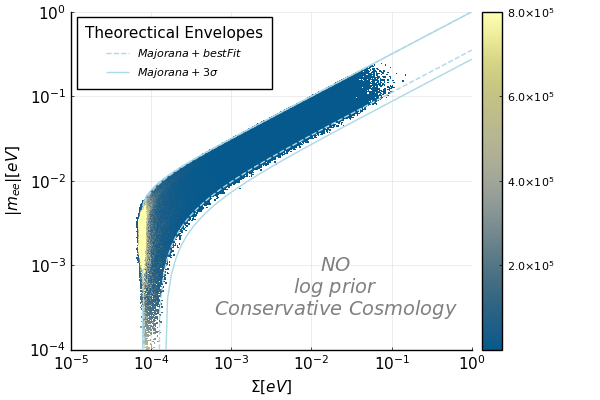}
\\
\includegraphics[width=7.2cm]{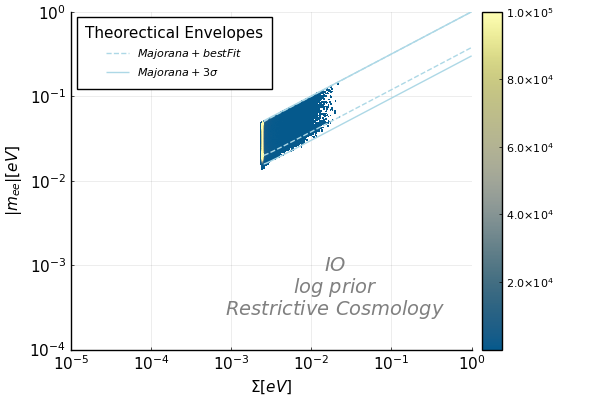} & \includegraphics[width=7.2cm]{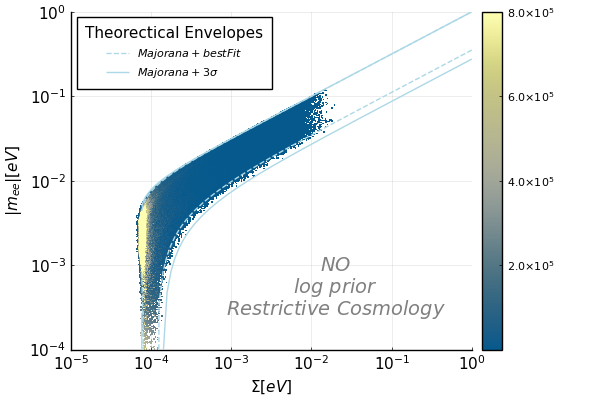}
\end{tabular}
\caption{\label{fig:KATRIN}Heat map of posterior probability density for both combinations of cosmological data sets and prior on $m_{\rm lightest}$. The upper panels depict the flat prior while the lower panels show the log prior. As for the plots in the main text, we used a random subset of the posterior sample with a size of $1 \times 10^6$. }
\end{figure*}

When considering laboratory-based experiments, there is a long history of attempts to measure the neutrino mass in single beta decay experiments. These types of experiments are looking for deviations in the kinematic endpoint of the resulting electron spectrum, and what they constrain (at least with realistic resolution) is the so-called \emph{effective electron neutrino mass} $m_\beta$, defined by\footnote{Note that, in fact, there is no mass that could be associated with an electron neutrino, as this is a not a mass eigenstate. Thus, the $m_\beta$ defined in Eq.~\eqref{eq:mbeta2} is really nothing more than an effective quantity, however, an observable one.}
\begin{equation}
 m_\beta^2 = m_1^2 c_{12}^2 c_{13}^2 + m_2^2 s_{12}^2 c_{13}^2 + m_3^2 s_{13}^2.
 \label{eq:mbeta2}
\end{equation}
Constraints on $m_\beta$ have mainly been derived from tritium decay, by both MAINZ~\cite{Kraus_2005} and Troitsk~\cite{LOBASHEV1999227}, however also future proposals focus on tritium\footnote{Alternative isotopes, such as rhenium as used by MARE~\cite{FERRI2015227}, do not seem to be competitive.} (such as Project~8~\cite{PhysRevD.80.051301}, which aims at detecting cyclotron radiation from relativistic tritium ions). The most prominent on-going effort is provided by KATRIN~\cite{KATRIN:2001ttj}, which is data taking and whose sensitivity on $m_\beta$ is of $\mathcal{O}(1~{\rm eV})$, with quoted values being dependent on the statistics used (e.g., Ref.~\cite{PhysRevD.76.113005} predicts a sensitivity of 0.20~eV based on frequentist methods and a sensitivity of 0.17~eV when using Bayesian statistics with a flat prior $m_\beta^2$).

The heat maps for our posterior probability densities for all combinations used of orderings and cosmological limits are displayed in Fig.~\ref{fig:KATRIN}. The underlying data are equivalent to those shown in the main text. As can be seen, the generic tendencies are basically the same as for the $\Sigma$--$|m_{ee}|$ plot. However, even if the ultimate sensitivity limit of KATRIN is reached, it will not add a significant constraint on the parameter space for the scenarios investigated. This is our justification for neglecting data from single beta decay experiments.

%%%%%%%%%%%%%%%%%%%%%%%%%%%%%%%%%%%%%%%%%%%%%%%%%%%%%%%%%%%%%%%%%%%%%%%%%%
\section*{Conclusion}
%%%%%%%%%%%%%%%%%%%%%%%%%%%%%%%%%%%%%%%%%%%%%%%%%%%%%%%%%%%%%%%%%%%%%%%%%%

The probability density heat maps and $0\nu\beta\beta$ results are based on currently available data. They should be updated as new data becomes available.
\onecolumngrid
%\end{widetext}

%\input{GlobalBayesNeutrino_Archive.bbl}
%\bibliography{GlobalBayesNeutrino_Archive.bbl}
\clearpage
\bibliography{main.bib}

\begin{thebibliography}{40}%
\makeatletter
\providecommand \@ifxundefined [1]{%
 \@ifx{#1\undefined}
}%
\providecommand \@ifnum [1]{%
 \ifnum #1\expandafter \@firstoftwo
 \else \expandafter \@secondoftwo
 \fi
}%
\providecommand \@ifx [1]{%
 \ifx #1\expandafter \@firstoftwo
 \else \expandafter \@secondoftwo
 \fi
}%
\providecommand \natexlab [1]{#1}%
\providecommand \enquote  [1]{``#1''}%
\providecommand \bibnamefont  [1]{#1}%
\providecommand \bibfnamefont [1]{#1}%
\providecommand \citenamefont [1]{#1}%
\providecommand \href@noop [0]{\@secondoftwo}%
\providecommand \href [0]{\begingroup \@sanitize@url \@href}%
\providecommand \@href[1]{\@@startlink{#1}\@@href}%
\providecommand \@@href[1]{\endgroup#1\@@endlink}%
\providecommand \@sanitize@url [0]{\catcode `\\12\catcode `\$12\catcode
  `\&12\catcode `\#12\catcode `\^12\catcode `\_12\catcode `\%12\relax}%
\providecommand \@@startlink[1]{}%
\providecommand \@@endlink[0]{}%
\providecommand \url  [0]{\begingroup\@sanitize@url \@url }%
\providecommand \@url [1]{\endgroup\@href {#1}{\urlprefix }}%
\providecommand \urlprefix  [0]{URL }%
\providecommand \Eprint [0]{\href }%
\providecommand \doibase [0]{http://dx.doi.org/}%
\providecommand \selectlanguage [0]{\@gobble}%
\providecommand \bibinfo  [0]{\@secondoftwo}%
\providecommand \bibfield  [0]{\@secondoftwo}%
\providecommand \translation [1]{[#1]}%
\providecommand \BibitemOpen [0]{}%
\providecommand \bibitemStop [0]{}%
\providecommand \bibitemNoStop [0]{.\EOS\space}%
\providecommand \EOS [0]{\spacefactor3000\relax}%
\providecommand \BibitemShut  [1]{\csname bibitem#1\endcsname}%
\let\auto@bib@innerbib\@empty
%</preamble>
\bibitem [{\citenamefont {Esteban}\ \emph {et~al.}(2017)\citenamefont
  {Esteban}, \citenamefont {Gonzalez-Garcia}, \citenamefont {Maltoni},
  \citenamefont {Martinez-Soler},\ and\ \citenamefont
  {Schwetz}}]{Esteban:2016qun}%
  \BibitemOpen
  \bibfield  {author} {\bibinfo {author} {\bibfnamefont {I.}~\bibnamefont
  {Esteban}}, \bibinfo {author} {\bibfnamefont {M.~C.}\ \bibnamefont
  {Gonzalez-Garcia}}, \bibinfo {author} {\bibfnamefont {M.}~\bibnamefont
  {Maltoni}}, \bibinfo {author} {\bibfnamefont {I.}~\bibnamefont
  {Martinez-Soler}}, \ and\ \bibinfo {author} {\bibfnamefont {T.}~\bibnamefont
  {Schwetz}},\ }\href {\doibase 10.1007/JHEP01(2017)087} {\bibfield  {journal}
  {\bibinfo  {journal} {JHEP}\ }\textbf {\bibinfo {volume} {01}},\ \bibinfo
  {pages} {087} (\bibinfo {year} {2017})},\ \Eprint
  {http://arxiv.org/abs/1611.01514} {arXiv:1611.01514 [hep-ph]} \BibitemShut
  {NoStop}%
%%CITATION = ARXIV:1611.01514;%%
\bibitem [{\citenamefont {Gando}\ \emph
  {et~al.}(2016{\natexlab{a}})\citenamefont {Gando} \emph
  {et~al.}}]{KamLAND-Zen:2016pfg}%
  \BibitemOpen
  \bibfield  {author} {\bibinfo {author} {\bibfnamefont {A.}~\bibnamefont
  {Gando}} \emph {et~al.} (\bibinfo {collaboration} {KamLAND-Zen}),\ }\href
  {\doibase 10.1103/PhysRevLett.117.109903, 10.1103/PhysRevLett.117.082503}
  {\bibfield  {journal} {\bibinfo  {journal} {Phys. Rev. Lett.}\ }\textbf
  {\bibinfo {volume} {117}},\ \bibinfo {pages} {082503} (\bibinfo {year}
  {2016}{\natexlab{a}})},\ \bibinfo {note} {[Addendum: Phys. Rev.
  Lett.117,no.10,109903(2016)]},\ \Eprint {http://arxiv.org/abs/1605.02889}
  {arXiv:1605.02889 [hep-ex]} \BibitemShut {NoStop}%
%%CITATION = ARXIV:1605.02889;%%
\bibitem [{\citenamefont {Albert}\ \emph
  {et~al.}(2014{\natexlab{a}})\citenamefont {Albert} \emph
  {et~al.}}]{Albert:2014awa}%
  \BibitemOpen
  \bibfield  {author} {\bibinfo {author} {\bibfnamefont {J.~B.}\ \bibnamefont
  {Albert}} \emph {et~al.} (\bibinfo {collaboration} {EXO-200}),\ }\href
  {\doibase 10.1038/nature13432} {\bibfield  {journal} {\bibinfo  {journal}
  {Nature}\ }\textbf {\bibinfo {volume} {510}},\ \bibinfo {pages} {229}
  (\bibinfo {year} {2014}{\natexlab{a}})},\ \Eprint
  {http://arxiv.org/abs/1402.6956} {arXiv:1402.6956 [nucl-ex]} \BibitemShut
  {NoStop}%
%%CITATION = ARXIV:1402.6956;%%
\bibitem [{\citenamefont {Agostini}\ \emph
  {et~al.}(2017{\natexlab{a}})\citenamefont {Agostini} \emph
  {et~al.}}]{Agostini:2017iyd}%
  \BibitemOpen
  \bibfield  {author} {\bibinfo {author} {\bibfnamefont {M.}~\bibnamefont
  {Agostini}} \emph {et~al.},\ }\href {\doibase 10.1038/nature21717} {\
  (\bibinfo {year} {2017}{\natexlab{a}}),\ 10.1038/nature21717},\ \Eprint
  {http://arxiv.org/abs/1703.00570} {arXiv:1703.00570 [nucl-ex]} \BibitemShut
  {NoStop}%
%%CITATION = ARXIV:1703.00570;%%
\bibitem [{\citenamefont {Gerbino}\ \emph {et~al.}(2016)\citenamefont
  {Gerbino}, \citenamefont {Lattanzi}, \citenamefont {Mena},\ and\
  \citenamefont {Freese}}]{Gerbino:2016ehw}%
  \BibitemOpen
  \bibfield  {author} {\bibinfo {author} {\bibfnamefont {M.}~\bibnamefont
  {Gerbino}}, \bibinfo {author} {\bibfnamefont {M.}~\bibnamefont {Lattanzi}},
  \bibinfo {author} {\bibfnamefont {O.}~\bibnamefont {Mena}}, \ and\ \bibinfo
  {author} {\bibfnamefont {K.}~\bibnamefont {Freese}},\ }\href@noop {} {\
  (\bibinfo {year} {2016})},\ \Eprint {http://arxiv.org/abs/1611.07847}
  {arXiv:1611.07847 [astro-ph.CO]} \BibitemShut {NoStop}%
%%CITATION = ARXIV:1611.07847;%%
\bibitem [{\citenamefont {Capozzi}\ \emph {et~al.}(2017)\citenamefont
  {Capozzi}, \citenamefont {Di~Valentino}, \citenamefont {Lisi}, \citenamefont
  {Marrone}, \citenamefont {Melchiorri},\ and\ \citenamefont
  {Palazzo}}]{Capozzi:2017ipn}%
  \BibitemOpen
  \bibfield  {author} {\bibinfo {author} {\bibfnamefont {F.}~\bibnamefont
  {Capozzi}}, \bibinfo {author} {\bibfnamefont {E.}~\bibnamefont
  {Di~Valentino}}, \bibinfo {author} {\bibfnamefont {E.}~\bibnamefont {Lisi}},
  \bibinfo {author} {\bibfnamefont {A.}~\bibnamefont {Marrone}}, \bibinfo
  {author} {\bibfnamefont {A.}~\bibnamefont {Melchiorri}}, \ and\ \bibinfo
  {author} {\bibfnamefont {A.}~\bibnamefont {Palazzo}},\ }\href@noop {} {\
  (\bibinfo {year} {2017})},\ \Eprint {http://arxiv.org/abs/1703.04471}
  {arXiv:1703.04471 [hep-ph]} \BibitemShut {NoStop}%
%%CITATION = ARXIV:1703.04471;%%
\bibitem [{\citenamefont {Simpson}\ \emph {et~al.}(2017)\citenamefont
  {Simpson}, \citenamefont {Jimenez}, \citenamefont {Pena-Garay},\ and\
  \citenamefont {Verde}}]{Simpson:2017qvj}%
  \BibitemOpen
  \bibfield  {author} {\bibinfo {author} {\bibfnamefont {F.}~\bibnamefont
  {Simpson}}, \bibinfo {author} {\bibfnamefont {R.}~\bibnamefont {Jimenez}},
  \bibinfo {author} {\bibfnamefont {C.}~\bibnamefont {Pena-Garay}}, \ and\
  \bibinfo {author} {\bibfnamefont {L.}~\bibnamefont {Verde}},\ }\href@noop {}
  {\  (\bibinfo {year} {2017})},\ \Eprint {http://arxiv.org/abs/1703.03425}
  {arXiv:1703.03425 [astro-ph.CO]} \BibitemShut {NoStop}%
%%CITATION = ARXIV:1703.03425;%%
\bibitem [{\citenamefont {Schwetz}\ \emph {et~al.}(2017)\citenamefont
  {Schwetz}, \citenamefont {Freese}, \citenamefont {Gerbino}, \citenamefont
  {Giusarma}, \citenamefont {Hannestad}, \citenamefont {Lattanzi},
  \citenamefont {Mena},\ and\ \citenamefont {Vagnozzi}}]{Schwetz:2017fey}%
  \BibitemOpen
  \bibfield  {author} {\bibinfo {author} {\bibfnamefont {T.}~\bibnamefont
  {Schwetz}}, \bibinfo {author} {\bibfnamefont {K.}~\bibnamefont {Freese}},
  \bibinfo {author} {\bibfnamefont {M.}~\bibnamefont {Gerbino}}, \bibinfo
  {author} {\bibfnamefont {E.}~\bibnamefont {Giusarma}}, \bibinfo {author}
  {\bibfnamefont {S.}~\bibnamefont {Hannestad}}, \bibinfo {author}
  {\bibfnamefont {M.}~\bibnamefont {Lattanzi}}, \bibinfo {author}
  {\bibfnamefont {O.}~\bibnamefont {Mena}}, \ and\ \bibinfo {author}
  {\bibfnamefont {S.}~\bibnamefont {Vagnozzi}},\ }\href@noop {} {\  (\bibinfo
  {year} {2017})},\ \Eprint {http://arxiv.org/abs/1703.04585} {arXiv:1703.04585
  [astro-ph.CO]} \BibitemShut {NoStop}%
%%CITATION = ARXIV:1703.04585;%%
\bibitem [{\citenamefont {Schervish}(1996)}]{Schervish:PValues}%
  \BibitemOpen
  \bibfield  {author} {\bibinfo {author} {\bibfnamefont {M.}~\bibnamefont
  {Schervish}},\ }\href {\doibase 10.1080/00031305.1996.10474380} {\bibfield
  {journal} {\bibinfo  {journal} {The American Statistician}\ }\textbf
  {\bibinfo {volume} {50}},\ \bibinfo {pages} {203} (\bibinfo {year} {1996})},\
  \Eprint
  {http://arxiv.org/abs/http://dx.doi.org/10.1080/00031305.1996.10474380}
  {http://dx.doi.org/10.1080/00031305.1996.10474380} \BibitemShut {NoStop}%
\bibitem [{\citenamefont {Neyman}(1977)}]{Neyman1977-NEYFPA}%
  \BibitemOpen
  \bibfield  {author} {\bibinfo {author} {\bibfnamefont {J.}~\bibnamefont
  {Neyman}},\ }\href@noop {} {\bibfield  {journal} {\bibinfo  {journal}
  {Synthese}\ }\textbf {\bibinfo {volume} {36}},\ \bibinfo {pages} {97}
  (\bibinfo {year} {1977})}\BibitemShut {NoStop}%
\bibitem [{\citenamefont {Davidson}\ \emph {et~al.}(2007)\citenamefont
  {Davidson}, \citenamefont {Isidori},\ and\ \citenamefont
  {Strumia}}]{Davidson:2006tg}%
  \BibitemOpen
  \bibfield  {author} {\bibinfo {author} {\bibfnamefont {S.}~\bibnamefont
  {Davidson}}, \bibinfo {author} {\bibfnamefont {G.}~\bibnamefont {Isidori}}, \
  and\ \bibinfo {author} {\bibfnamefont {A.}~\bibnamefont {Strumia}},\ }\href
  {\doibase 10.1016/j.physletb.2007.01.015} {\bibfield  {journal} {\bibinfo
  {journal} {Phys. Lett.}\ }\textbf {\bibinfo {volume} {B646}},\ \bibinfo
  {pages} {100} (\bibinfo {year} {2007})},\ \Eprint
  {http://arxiv.org/abs/hep-ph/0611389} {arXiv:hep-ph/0611389 [hep-ph]}
  \BibitemShut {NoStop}%
%%CITATION = HEP-PH/0611389;%%
\bibitem [{\citenamefont {Schulz}\ \emph {et~al.}(2021)\citenamefont {Schulz},
  \citenamefont {Beaujean}, \citenamefont {Caldwell}, \citenamefont {Grunwald},
  \citenamefont {Hafych}, \citenamefont {Kr{\"o}ninger}, \citenamefont
  {Cagnina}, \citenamefont {R{\"o}hrig},\ and\ \citenamefont
  {Shtembari}}]{Schulz:2021BAT}%
  \BibitemOpen
  \bibfield  {author} {\bibinfo {author} {\bibfnamefont {O.}~\bibnamefont
  {Schulz}}, \bibinfo {author} {\bibfnamefont {F.}~\bibnamefont {Beaujean}},
  \bibinfo {author} {\bibfnamefont {A.}~\bibnamefont {Caldwell}}, \bibinfo
  {author} {\bibfnamefont {C.}~\bibnamefont {Grunwald}}, \bibinfo {author}
  {\bibfnamefont {V.}~\bibnamefont {Hafych}}, \bibinfo {author} {\bibfnamefont
  {K.}~\bibnamefont {Kr{\"o}ninger}}, \bibinfo {author} {\bibfnamefont {S.~L.}\
  \bibnamefont {Cagnina}}, \bibinfo {author} {\bibfnamefont {L.}~\bibnamefont
  {R{\"o}hrig}}, \ and\ \bibinfo {author} {\bibfnamefont {L.}~\bibnamefont
  {Shtembari}},\ }\href {\doibase 10.1007/s42979-021-00626-4} {\bibfield
  {journal} {\bibinfo  {journal} {SN Computer Science}\ }\textbf {\bibinfo
  {volume} {2}},\ \bibinfo {pages} {210} (\bibinfo {year} {2021})}\BibitemShut
  {NoStop}%
\bibitem [{\citenamefont {Lindner}\ \emph {et~al.}(2006)\citenamefont
  {Lindner}, \citenamefont {Merle},\ and\ \citenamefont
  {Rodejohann}}]{Lindner:2005kr}%
  \BibitemOpen
  \bibfield  {author} {\bibinfo {author} {\bibfnamefont {M.}~\bibnamefont
  {Lindner}}, \bibinfo {author} {\bibfnamefont {A.}~\bibnamefont {Merle}}, \
  and\ \bibinfo {author} {\bibfnamefont {W.}~\bibnamefont {Rodejohann}},\
  }\href {\doibase 10.1103/PhysRevD.73.053005} {\bibfield  {journal} {\bibinfo
  {journal} {Phys. Rev.}\ }\textbf {\bibinfo {volume} {D73}},\ \bibinfo {pages}
  {053005} (\bibinfo {year} {2006})},\ \Eprint
  {http://arxiv.org/abs/hep-ph/0512143} {arXiv:hep-ph/0512143 [hep-ph]}
  \BibitemShut {NoStop}%
%%CITATION = HEP-PH/0512143;%%
\bibitem [{\citenamefont {Merle}\ and\ \citenamefont
  {Rodejohann}(2006)}]{Merle:2006du}%
  \BibitemOpen
  \bibfield  {author} {\bibinfo {author} {\bibfnamefont {A.}~\bibnamefont
  {Merle}}\ and\ \bibinfo {author} {\bibfnamefont {W.}~\bibnamefont
  {Rodejohann}},\ }\href {\doibase 10.1103/PhysRevD.73.073012} {\bibfield
  {journal} {\bibinfo  {journal} {Phys. Rev.}\ }\textbf {\bibinfo {volume}
  {D73}},\ \bibinfo {pages} {073012} (\bibinfo {year} {2006})},\ \Eprint
  {http://arxiv.org/abs/hep-ph/0603111} {arXiv:hep-ph/0603111 [hep-ph]}
  \BibitemShut {NoStop}%
%%CITATION = HEP-PH/0603111;%%
\bibitem [{\citenamefont {Maneschg}\ \emph {et~al.}(2009)\citenamefont
  {Maneschg}, \citenamefont {Merle},\ and\ \citenamefont
  {Rodejohann}}]{Maneschg:2008sf}%
  \BibitemOpen
  \bibfield  {author} {\bibinfo {author} {\bibfnamefont {W.}~\bibnamefont
  {Maneschg}}, \bibinfo {author} {\bibfnamefont {A.}~\bibnamefont {Merle}}, \
  and\ \bibinfo {author} {\bibfnamefont {W.}~\bibnamefont {Rodejohann}},\
  }\href {\doibase 10.1209/0295-5075/85/51002} {\bibfield  {journal} {\bibinfo
  {journal} {Europhys. Lett.}\ }\textbf {\bibinfo {volume} {85}},\ \bibinfo
  {pages} {51002} (\bibinfo {year} {2009})},\ \Eprint
  {http://arxiv.org/abs/0812.0479} {arXiv:0812.0479 [hep-ph]} \BibitemShut
  {NoStop}%
%%CITATION = ARXIV:0812.0479;%%
\bibitem [{\citenamefont {Hahn}(2005)}]{HAHN200578}%
  \BibitemOpen
  \bibfield  {author} {\bibinfo {author} {\bibfnamefont {T.}~\bibnamefont
  {Hahn}},\ }\href {\doibase https://doi.org/10.1016/j.cpc.2005.01.010}
  {\bibfield  {journal} {\bibinfo  {journal} {Computer Physics Communications}\
  }\textbf {\bibinfo {volume} {168}},\ \bibinfo {pages} {78} (\bibinfo {year}
  {2005})}\BibitemShut {NoStop}%
\bibitem [{\citenamefont {Doi}\ \emph {et~al.}(1985)\citenamefont {Doi},
  \citenamefont {Kotani},\ and\ \citenamefont {Takasugi}}]{Doi:1985dx}%
  \BibitemOpen
  \bibfield  {author} {\bibinfo {author} {\bibfnamefont {M.}~\bibnamefont
  {Doi}}, \bibinfo {author} {\bibfnamefont {T.}~\bibnamefont {Kotani}}, \ and\
  \bibinfo {author} {\bibfnamefont {E.}~\bibnamefont {Takasugi}},\ }\href
  {\doibase 10.1143/PTPS.83.1} {\bibfield  {journal} {\bibinfo  {journal}
  {Prog. Theor. Phys. Suppl.}\ }\textbf {\bibinfo {volume} {83}},\ \bibinfo
  {pages} {1} (\bibinfo {year} {1985})}\BibitemShut {NoStop}%
%%CITATION = PTPSA,83,1;%%
\bibitem [{\citenamefont {Suhonen}\ and\ \citenamefont
  {Civitarese}(1998)}]{Suhonen:1998ck}%
  \BibitemOpen
  \bibfield  {author} {\bibinfo {author} {\bibfnamefont {J.}~\bibnamefont
  {Suhonen}}\ and\ \bibinfo {author} {\bibfnamefont {O.}~\bibnamefont
  {Civitarese}},\ }\href {\doibase 10.1016/S0370-1573(97)00087-2} {\bibfield
  {journal} {\bibinfo  {journal} {Phys. Rept.}\ }\textbf {\bibinfo {volume}
  {300}},\ \bibinfo {pages} {123} (\bibinfo {year} {1998})}\BibitemShut
  {NoStop}%
%%CITATION = PRPLC,300,123;%%
\bibitem [{\citenamefont {Rodejohann}(2011)}]{Rodejohann:2011mu}%
  \BibitemOpen
  \bibfield  {author} {\bibinfo {author} {\bibfnamefont {W.}~\bibnamefont
  {Rodejohann}},\ }\href {\doibase 10.1142/S0218301311020186} {\bibfield
  {journal} {\bibinfo  {journal} {Int. J. Mod. Phys.}\ }\textbf {\bibinfo
  {volume} {E20}},\ \bibinfo {pages} {1833} (\bibinfo {year} {2011})},\ \Eprint
  {http://arxiv.org/abs/1106.1334} {arXiv:1106.1334 [hep-ph]} \BibitemShut
  {NoStop}%
%%CITATION = ARXIV:1106.1334;%%
\bibitem [{\citenamefont {Menendez}\ \emph {et~al.}(2009)\citenamefont
  {Menendez}, \citenamefont {Poves}, \citenamefont {Caurier},\ and\
  \citenamefont {Nowacki}}]{Menendez:2008jp}%
  \BibitemOpen
  \bibfield  {author} {\bibinfo {author} {\bibfnamefont {J.}~\bibnamefont
  {Menendez}}, \bibinfo {author} {\bibfnamefont {A.}~\bibnamefont {Poves}},
  \bibinfo {author} {\bibfnamefont {E.}~\bibnamefont {Caurier}}, \ and\
  \bibinfo {author} {\bibfnamefont {F.}~\bibnamefont {Nowacki}},\ }\href
  {\doibase 10.1016/j.nuclphysa.2008.12.005} {\bibfield  {journal} {\bibinfo
  {journal} {Nucl. Phys.}\ }\textbf {\bibinfo {volume} {A818}},\ \bibinfo
  {pages} {139} (\bibinfo {year} {2009})},\ \Eprint
  {http://arxiv.org/abs/0801.3760} {arXiv:0801.3760 [nucl-th]} \BibitemShut
  {NoStop}%
%%CITATION = ARXIV:0801.3760;%%
\bibitem [{\citenamefont {Simkovic}\ \emph {et~al.}(2013)\citenamefont
  {Simkovic}, \citenamefont {Rodin}, \citenamefont {Faessler},\ and\
  \citenamefont {Vogel}}]{Simkovic:2013qiy}%
  \BibitemOpen
  \bibfield  {author} {\bibinfo {author} {\bibfnamefont {F.}~\bibnamefont
  {Simkovic}}, \bibinfo {author} {\bibfnamefont {V.}~\bibnamefont {Rodin}},
  \bibinfo {author} {\bibfnamefont {A.}~\bibnamefont {Faessler}}, \ and\
  \bibinfo {author} {\bibfnamefont {P.}~\bibnamefont {Vogel}},\ }\href
  {\doibase 10.1103/PhysRevC.87.045501} {\bibfield  {journal} {\bibinfo
  {journal} {Phys. Rev.}\ }\textbf {\bibinfo {volume} {C87}},\ \bibinfo {pages}
  {045501} (\bibinfo {year} {2013})},\ \Eprint {http://arxiv.org/abs/1302.1509}
  {arXiv:1302.1509 [nucl-th]} \BibitemShut {NoStop}%
%%CITATION = ARXIV:1302.1509;%%
\bibitem [{\citenamefont {Meroni}\ \emph {et~al.}(2013)\citenamefont {Meroni},
  \citenamefont {Petcov},\ and\ \citenamefont {Simkovic}}]{Meroni:2012qf}%
  \BibitemOpen
  \bibfield  {author} {\bibinfo {author} {\bibfnamefont {A.}~\bibnamefont
  {Meroni}}, \bibinfo {author} {\bibfnamefont {S.~T.}\ \bibnamefont {Petcov}},
  \ and\ \bibinfo {author} {\bibfnamefont {F.}~\bibnamefont {Simkovic}},\
  }\href {\doibase 10.1007/JHEP02(2013)025} {\bibfield  {journal} {\bibinfo
  {journal} {JHEP}\ }\textbf {\bibinfo {volume} {1302}},\ \bibinfo {pages}
  {025} (\bibinfo {year} {2013})},\ \Eprint {http://arxiv.org/abs/1212.1331}
  {arXiv:1212.1331 [hep-ph]} \BibitemShut {NoStop}%
%%CITATION = ARXIV:1212.1331;%%
\bibitem [{\citenamefont {Bhupal~Dev}\ \emph {et~al.}(2013)\citenamefont
  {Bhupal~Dev}, \citenamefont {Goswami}, \citenamefont {Mitra},\ and\
  \citenamefont {Rodejohann}}]{Dev:2013vxa}%
  \BibitemOpen
  \bibfield  {author} {\bibinfo {author} {\bibfnamefont {P.~S.}\ \bibnamefont
  {Bhupal~Dev}}, \bibinfo {author} {\bibfnamefont {S.}~\bibnamefont {Goswami}},
  \bibinfo {author} {\bibfnamefont {M.}~\bibnamefont {Mitra}}, \ and\ \bibinfo
  {author} {\bibfnamefont {W.}~\bibnamefont {Rodejohann}},\ }\href@noop {} {\
  (\bibinfo {year} {2013})},\ \Eprint {http://arxiv.org/abs/1305.0056}
  {arXiv:1305.0056 [hep-ph]} \BibitemShut {NoStop}%
%%CITATION = ARXIV:1305.0056;%%
\bibitem [{\citenamefont {Rodriguez}\ and\ \citenamefont
  {Martinez-Pinedo}(2010)}]{Rodriguez:2010mn}%
  \BibitemOpen
  \bibfield  {author} {\bibinfo {author} {\bibfnamefont {T.~R.}\ \bibnamefont
  {Rodriguez}}\ and\ \bibinfo {author} {\bibfnamefont {G.}~\bibnamefont
  {Martinez-Pinedo}},\ }\href {\doibase 10.1103/PhysRevLett.105.252503}
  {\bibfield  {journal} {\bibinfo  {journal} {Phys. Rev. Lett.}\ }\textbf
  {\bibinfo {volume} {105}},\ \bibinfo {pages} {252503} (\bibinfo {year}
  {2010})},\ \Eprint {http://arxiv.org/abs/1008.5260} {arXiv:1008.5260
  [nucl-th]} \BibitemShut {NoStop}%
%%CITATION = ARXIV:1008.5260;%%
\bibitem [{\citenamefont {Barea}\ \emph {et~al.}(2013)\citenamefont {Barea},
  \citenamefont {Kotila},\ and\ \citenamefont {Iachello}}]{Barea:2013bz}%
  \BibitemOpen
  \bibfield  {author} {\bibinfo {author} {\bibfnamefont {J.}~\bibnamefont
  {Barea}}, \bibinfo {author} {\bibfnamefont {J.}~\bibnamefont {Kotila}}, \
  and\ \bibinfo {author} {\bibfnamefont {F.}~\bibnamefont {Iachello}},\ }\href
  {\doibase 10.1103/PhysRevC.87.014315} {\bibfield  {journal} {\bibinfo
  {journal} {Phys. Rev.}\ }\textbf {\bibinfo {volume} {C87}},\ \bibinfo {pages}
  {014315} (\bibinfo {year} {2013})},\ \Eprint {http://arxiv.org/abs/1301.4203}
  {arXiv:1301.4203 [nucl-th]} \BibitemShut {NoStop}%
%%CITATION = ARXIV:1301.4203;%%
\bibitem [{\citenamefont {Suhonen}\ and\ \citenamefont
  {Civitarese}(2010)}]{Suhonen:2010zzc}%
  \BibitemOpen
  \bibfield  {author} {\bibinfo {author} {\bibfnamefont {J.}~\bibnamefont
  {Suhonen}}\ and\ \bibinfo {author} {\bibfnamefont {O.}~\bibnamefont
  {Civitarese}},\ }\href {\doibase 10.1016/j.nuclphysa.2010.08.003} {\bibfield
  {journal} {\bibinfo  {journal} {Nucl. Phys.}\ }\textbf {\bibinfo {volume}
  {A847}},\ \bibinfo {pages} {207} (\bibinfo {year} {2010})}\BibitemShut
  {NoStop}%
%%CITATION = NUPHA,A847,207;%%
\bibitem [{\citenamefont {Mustonen}\ and\ \citenamefont
  {Engel}(2013)}]{Mustonen:2013zu}%
  \BibitemOpen
  \bibfield  {author} {\bibinfo {author} {\bibfnamefont {M.~T.}\ \bibnamefont
  {Mustonen}}\ and\ \bibinfo {author} {\bibfnamefont {J.}~\bibnamefont
  {Engel}},\ }\href@noop {} {\  (\bibinfo {year} {2013})},\ \Eprint
  {http://arxiv.org/abs/1301.6997} {arXiv:1301.6997 [nucl-th]} \BibitemShut
  {NoStop}%
%%CITATION = ARXIV:1301.6997;%%
\bibitem [{\citenamefont {Agostini}\ \emph
  {et~al.}(2017{\natexlab{b}})\citenamefont {Agostini} \emph
  {et~al.}}]{ref:GERDA}%
  \BibitemOpen
  \bibfield  {author} {\bibinfo {author} {\bibfnamefont {M.}~\bibnamefont
  {Agostini}} \emph {et~al.},\ }\href {\doibase 10.1038/nature21717} {\
  (\bibinfo {year} {2017}{\natexlab{b}}),\ 10.1038/nature21717},\ \Eprint
  {http://arxiv.org/abs/1703.00570} {arXiv:1703.00570 [nucl-ex]} \BibitemShut
  {NoStop}%
%%CITATION = ARXIV:1703.00570;%%
\bibitem [{\citenamefont {Gando}\ \emph
  {et~al.}(2016{\natexlab{b}})\citenamefont {Gando} \emph
  {et~al.}}]{ref:Kamland-Zen}%
  \BibitemOpen
  \bibfield  {author} {\bibinfo {author} {\bibfnamefont {A.}~\bibnamefont
  {Gando}} \emph {et~al.} (\bibinfo {collaboration} {KamLAND-Zen}),\ }\href
  {\doibase 10.1103/PhysRevLett.117.109903, 10.1103/PhysRevLett.117.082503}
  {\bibfield  {journal} {\bibinfo  {journal} {Phys. Rev. Lett.}\ }\textbf
  {\bibinfo {volume} {117}},\ \bibinfo {pages} {082503} (\bibinfo {year}
  {2016}{\natexlab{b}})},\ \bibinfo {note} {[Addendum: Phys. Rev.
  Lett.117,no.10,109903(2016)]},\ \Eprint {http://arxiv.org/abs/1605.02889}
  {arXiv:1605.02889 [hep-ex]} \BibitemShut {NoStop}%
%%CITATION = ARXIV:1605.02889;%%
\bibitem [{\citenamefont {Albert}\ \emph
  {et~al.}(2014{\natexlab{b}})\citenamefont {Albert} \emph {et~al.}}]{ref:EXO}%
  \BibitemOpen
  \bibfield  {author} {\bibinfo {author} {\bibfnamefont {J.~B.}\ \bibnamefont
  {Albert}} \emph {et~al.} (\bibinfo {collaboration} {EXO-200}),\ }\href
  {\doibase 10.1038/nature13432} {\bibfield  {journal} {\bibinfo  {journal}
  {Nature}\ }\textbf {\bibinfo {volume} {510}},\ \bibinfo {pages} {229}
  (\bibinfo {year} {2014}{\natexlab{b}})},\ \Eprint
  {http://arxiv.org/abs/1402.6956} {arXiv:1402.6956 [nucl-ex]} \BibitemShut
  {NoStop}%
%%CITATION = ARXIV:1402.6956;%%
\bibitem [{\citenamefont {King}\ \emph {et~al.}(2013)\citenamefont {King},
  \citenamefont {Merle},\ and\ \citenamefont {Stuart}}]{King_2013}%
  \BibitemOpen
  \bibfield  {author} {\bibinfo {author} {\bibfnamefont {S.~F.}\ \bibnamefont
  {King}}, \bibinfo {author} {\bibfnamefont {A.}~\bibnamefont {Merle}}, \ and\
  \bibinfo {author} {\bibfnamefont {A.~J.}\ \bibnamefont {Stuart}},\ }\href
  {\doibase 10.1007/jhep12(2013)005} {\bibfield  {journal} {\bibinfo  {journal}
  {Journal of High Energy Physics}\ }\textbf {\bibinfo {volume} {2013}}
  (\bibinfo {year} {2013}),\ 10.1007/jhep12(2013)005}\BibitemShut {NoStop}%
\bibitem [{\citenamefont {Gehrlein}\ \emph {et~al.}(2016)\citenamefont
  {Gehrlein}, \citenamefont {Merle},\ and\ \citenamefont
  {Spinrath}}]{PhysRevD.94.093003}%
  \BibitemOpen
  \bibfield  {author} {\bibinfo {author} {\bibfnamefont {J.}~\bibnamefont
  {Gehrlein}}, \bibinfo {author} {\bibfnamefont {A.}~\bibnamefont {Merle}}, \
  and\ \bibinfo {author} {\bibfnamefont {M.}~\bibnamefont {Spinrath}},\ }\href
  {\doibase 10.1103/PhysRevD.94.093003} {\bibfield  {journal} {\bibinfo
  {journal} {Phys. Rev. D}\ }\textbf {\bibinfo {volume} {94}},\ \bibinfo
  {pages} {093003} (\bibinfo {year} {2016})}\BibitemShut {NoStop}%
\bibitem [{\citenamefont {Gehrlein}\ \emph {et~al.}(2015)\citenamefont
  {Gehrlein}, \citenamefont {Merle},\ and\ \citenamefont
  {Spinrath}}]{Gehrlein_2015}%
  \BibitemOpen
  \bibfield  {author} {\bibinfo {author} {\bibfnamefont {J.}~\bibnamefont
  {Gehrlein}}, \bibinfo {author} {\bibfnamefont {A.}~\bibnamefont {Merle}}, \
  and\ \bibinfo {author} {\bibfnamefont {M.}~\bibnamefont {Spinrath}},\ }\href
  {\doibase 10.1007/jhep09(2015)066} {\bibfield  {journal} {\bibinfo  {journal}
  {Journal of High Energy Physics}\ }\textbf {\bibinfo {volume} {2015}}
  (\bibinfo {year} {2015}),\ 10.1007/jhep09(2015)066}\BibitemShut {NoStop}%
\bibitem [{\citenamefont {Agostini}\ \emph {et~al.}(2016)\citenamefont
  {Agostini}, \citenamefont {Merle},\ and\ \citenamefont
  {Zuber}}]{Agostini_2016}%
  \BibitemOpen
  \bibfield  {author} {\bibinfo {author} {\bibfnamefont {M.}~\bibnamefont
  {Agostini}}, \bibinfo {author} {\bibfnamefont {A.}~\bibnamefont {Merle}}, \
  and\ \bibinfo {author} {\bibfnamefont {K.}~\bibnamefont {Zuber}},\ }\href
  {\doibase 10.1140/epjc/s10052-016-4011-2} {\bibfield  {journal} {\bibinfo
  {journal} {The European Physical Journal C}\ }\textbf {\bibinfo {volume}
  {76}} (\bibinfo {year} {2016}),\ 10.1140/epjc/s10052-016-4011-2}\BibitemShut
  {NoStop}%
\bibitem [{\citenamefont {Kraus}\ \emph {et~al.}(2005)\citenamefont {Kraus},
  \citenamefont {Bornschein}, \citenamefont {Bornschein}, \citenamefont {Bonn},
  \citenamefont {Flatt}, \citenamefont {Kovalik}, \citenamefont {Ostrick},
  \citenamefont {Otten}, \citenamefont {Schall}, \citenamefont {Thümmler},\
  and\ \citenamefont {et~al.}}]{Kraus_2005}%
  \BibitemOpen
  \bibfield  {author} {\bibinfo {author} {\bibfnamefont {C.}~\bibnamefont
  {Kraus}}, \bibinfo {author} {\bibfnamefont {B.}~\bibnamefont {Bornschein}},
  \bibinfo {author} {\bibfnamefont {L.}~\bibnamefont {Bornschein}}, \bibinfo
  {author} {\bibfnamefont {J.}~\bibnamefont {Bonn}}, \bibinfo {author}
  {\bibfnamefont {B.}~\bibnamefont {Flatt}}, \bibinfo {author} {\bibfnamefont
  {A.}~\bibnamefont {Kovalik}}, \bibinfo {author} {\bibfnamefont
  {B.}~\bibnamefont {Ostrick}}, \bibinfo {author} {\bibfnamefont {E.~W.}\
  \bibnamefont {Otten}}, \bibinfo {author} {\bibfnamefont {J.~P.}\ \bibnamefont
  {Schall}}, \bibinfo {author} {\bibfnamefont {T.}~\bibnamefont {Thümmler}}, \
  and\ \bibinfo {author} {\bibnamefont {et~al.}},\ }\href {\doibase
  10.1140/epjc/s2005-02139-7} {\bibfield  {journal} {\bibinfo  {journal} {The
  European Physical Journal C}\ }\textbf {\bibinfo {volume} {40}},\ \bibinfo
  {pages} {447–468} (\bibinfo {year} {2005})}\BibitemShut {NoStop}%
\bibitem [{\citenamefont {Lobashev}\ \emph {et~al.}(1999)\citenamefont
  {Lobashev}, \citenamefont {Aseev}, \citenamefont {Belesev}, \citenamefont
  {Berlev}, \citenamefont {Geraskin}, \citenamefont {Golubev}, \citenamefont
  {Kazachenko}, \citenamefont {Kuznetsov}, \citenamefont {Ostroumov},
  \citenamefont {Rivkis}, \citenamefont {Stern}, \citenamefont {Titov},
  \citenamefont {Zadorozhny},\ and\ \citenamefont
  {Zakharov}}]{LOBASHEV1999227}%
  \BibitemOpen
  \bibfield  {author} {\bibinfo {author} {\bibfnamefont {V.}~\bibnamefont
  {Lobashev}}, \bibinfo {author} {\bibfnamefont {V.}~\bibnamefont {Aseev}},
  \bibinfo {author} {\bibfnamefont {A.}~\bibnamefont {Belesev}}, \bibinfo
  {author} {\bibfnamefont {A.}~\bibnamefont {Berlev}}, \bibinfo {author}
  {\bibfnamefont {E.}~\bibnamefont {Geraskin}}, \bibinfo {author}
  {\bibfnamefont {A.}~\bibnamefont {Golubev}}, \bibinfo {author} {\bibfnamefont
  {O.}~\bibnamefont {Kazachenko}}, \bibinfo {author} {\bibfnamefont
  {Y.}~\bibnamefont {Kuznetsov}}, \bibinfo {author} {\bibfnamefont
  {R.}~\bibnamefont {Ostroumov}}, \bibinfo {author} {\bibfnamefont
  {L.}~\bibnamefont {Rivkis}}, \bibinfo {author} {\bibfnamefont
  {B.}~\bibnamefont {Stern}}, \bibinfo {author} {\bibfnamefont
  {N.}~\bibnamefont {Titov}}, \bibinfo {author} {\bibfnamefont
  {S.}~\bibnamefont {Zadorozhny}}, \ and\ \bibinfo {author} {\bibfnamefont
  {Y.}~\bibnamefont {Zakharov}},\ }\href {\doibase
  https://doi.org/10.1016/S0370-2693(99)00781-9} {\bibfield  {journal}
  {\bibinfo  {journal} {Physics Letters B}\ }\textbf {\bibinfo {volume}
  {460}},\ \bibinfo {pages} {227} (\bibinfo {year} {1999})}\BibitemShut
  {NoStop}%
\bibitem [{\citenamefont {Ferri}\ \emph {et~al.}(2015)\citenamefont {Ferri},
  \citenamefont {Bagliani}, \citenamefont {Biasotti}, \citenamefont {Ceruti},
  \citenamefont {Corsini}, \citenamefont {Faverzani}, \citenamefont {Gatti},
  \citenamefont {Giachero}, \citenamefont {Gotti}, \citenamefont {Kilbourne},
  \citenamefont {Kling}, \citenamefont {Maino}, \citenamefont {Manfrinetti},
  \citenamefont {Nucciotti}, \citenamefont {Pessina}, \citenamefont
  {Pizzigoni}, \citenamefont {Gomes},\ and\ \citenamefont
  {Sisti}}]{FERRI2015227}%
  \BibitemOpen
  \bibfield  {author} {\bibinfo {author} {\bibfnamefont {E.}~\bibnamefont
  {Ferri}}, \bibinfo {author} {\bibfnamefont {D.}~\bibnamefont {Bagliani}},
  \bibinfo {author} {\bibfnamefont {M.}~\bibnamefont {Biasotti}}, \bibinfo
  {author} {\bibfnamefont {G.}~\bibnamefont {Ceruti}}, \bibinfo {author}
  {\bibfnamefont {D.}~\bibnamefont {Corsini}}, \bibinfo {author} {\bibfnamefont
  {M.}~\bibnamefont {Faverzani}}, \bibinfo {author} {\bibfnamefont
  {F.}~\bibnamefont {Gatti}}, \bibinfo {author} {\bibfnamefont
  {A.}~\bibnamefont {Giachero}}, \bibinfo {author} {\bibfnamefont
  {C.}~\bibnamefont {Gotti}}, \bibinfo {author} {\bibfnamefont
  {C.}~\bibnamefont {Kilbourne}}, \bibinfo {author} {\bibfnamefont
  {A.}~\bibnamefont {Kling}}, \bibinfo {author} {\bibfnamefont
  {M.}~\bibnamefont {Maino}}, \bibinfo {author} {\bibfnamefont
  {P.}~\bibnamefont {Manfrinetti}}, \bibinfo {author} {\bibfnamefont
  {A.}~\bibnamefont {Nucciotti}}, \bibinfo {author} {\bibfnamefont
  {G.}~\bibnamefont {Pessina}}, \bibinfo {author} {\bibfnamefont
  {G.}~\bibnamefont {Pizzigoni}}, \bibinfo {author} {\bibfnamefont {M.~R.}\
  \bibnamefont {Gomes}}, \ and\ \bibinfo {author} {\bibfnamefont
  {M.}~\bibnamefont {Sisti}},\ }\href {\doibase
  https://doi.org/10.1016/j.phpro.2014.12.037} {\bibfield  {journal} {\bibinfo
  {journal} {Physics Procedia}\ }\textbf {\bibinfo {volume} {61}},\ \bibinfo
  {pages} {227} (\bibinfo {year} {2015})},\ \bibinfo {note} {13th International
  Conference on Topics in Astroparticle and Underground Physics, TAUP
  2013}\BibitemShut {NoStop}%
\bibitem [{\citenamefont {Monreal}\ and\ \citenamefont
  {Formaggio}(2009)}]{PhysRevD.80.051301}%
  \BibitemOpen
  \bibfield  {author} {\bibinfo {author} {\bibfnamefont {B.}~\bibnamefont
  {Monreal}}\ and\ \bibinfo {author} {\bibfnamefont {J.~A.}\ \bibnamefont
  {Formaggio}},\ }\href {\doibase 10.1103/PhysRevD.80.051301} {\bibfield
  {journal} {\bibinfo  {journal} {Phys. Rev. D}\ }\textbf {\bibinfo {volume}
  {80}},\ \bibinfo {pages} {051301} (\bibinfo {year} {2009})}\BibitemShut
  {NoStop}%
\bibitem [{\citenamefont {Osipowicz}\ \emph {et~al.}(2001)\citenamefont
  {Osipowicz} \emph {et~al.}}]{KATRIN:2001ttj}%
  \BibitemOpen
  \bibfield  {author} {\bibinfo {author} {\bibfnamefont {A.}~\bibnamefont
  {Osipowicz}} \emph {et~al.} (\bibinfo {collaboration} {KATRIN}),\ }\href@noop
  {} {\  (\bibinfo {year} {2001})},\ \Eprint
  {http://arxiv.org/abs/hep-ex/0109033} {arXiv:hep-ex/0109033} \BibitemShut
  {NoStop}%
\bibitem [{\citenamefont {Host}\ \emph {et~al.}(2007)\citenamefont {Host},
  \citenamefont {Lahav}, \citenamefont {Abdalla},\ and\ \citenamefont
  {Eitel}}]{PhysRevD.76.113005}%
  \BibitemOpen
  \bibfield  {author} {\bibinfo {author} {\bibfnamefont {O.}~\bibnamefont
  {Host}}, \bibinfo {author} {\bibfnamefont {O.}~\bibnamefont {Lahav}},
  \bibinfo {author} {\bibfnamefont {F.~B.}\ \bibnamefont {Abdalla}}, \ and\
  \bibinfo {author} {\bibfnamefont {K.}~\bibnamefont {Eitel}},\ }\href
  {\doibase 10.1103/PhysRevD.76.113005} {\bibfield  {journal} {\bibinfo
  {journal} {Phys. Rev. D}\ }\textbf {\bibinfo {volume} {76}},\ \bibinfo
  {pages} {113005} (\bibinfo {year} {2007})}\BibitemShut {NoStop}%
\end{thebibliography}%
%%%%%%%%%%%%%%%%%%%%%%%%%%%%%%%%%%%%%%%%%%%%%%%%%%%%%%%%%%%%%%%%%%%%%%%%%%%%%%%%%%%%%%%%%%%%%%%%%%%%%%%%%%%%%%%%%%%%%%%%%%%%%%%%%%%%%%%%%%%%%%%%%%%%%%%%%%%%%%%%%%%%%%%%%%%%%%%%%%%%%%%%%%%%%%%%%%%%%%%%%%%%%%%%%%%%%%%%%%%%%%%%%%%%%%%%%%%%%
%%%%%%%%%%%%%%%%%%%%%%%%%%%%%%%%%%%%%%%%%%%%%%%%%%%%%%%%%%%%%%%%%%%%%%%%%%%%%%%%%%%%%%%%%%%%%%%%%%%%%%%%%%%%%%%%%%%%%%%%%%%%%%%%%%%%%%%%%%%%%%%%%%%%%%%%%%%%%%%%%%%%%%%%%%%%%%%%%%%%%%%%%%%%%%%%%%%%%%%%%%%%%%%%%%%%%%%%%%%%%%%%%%%%%%%%%%%%%
%%%%%%%%%%%%%%%%%%%%%%%%%%%%%%%%%%%%%%%%%%%%%%%%%%%%%%%%%%%%%%%%%%%%%%%%%%%%%%%%%%%%%%%%%%%%%%%%%%%%%%%%%%%%%%%%%%%%%%%%%%%%%%%%%%%%%%%%%%%%%%%%%%%%%%%%%%%%%%%%%%%%%%%%%%%%%%%%%%%%%%%%%%%%%%%%%%%%%%%%%%%%%%%%%%%%%%%%%%%%%%%%%%%%%%%%%%%%%

\end{document}